\documentclass[11pt]{article}

\usepackage[top=1.0in,right=1.0in,left=1.0in,bottom=1.75in]{geometry}
\usepackage{latexsym}
\usepackage{amssymb,amsbsy}
\usepackage{bm}
\usepackage{amsmath}
\usepackage{graphics,graphicx}
\usepackage{color}
\usepackage[affil-it,auth-sc]{authblk}
\usepackage{subcaption}
\usepackage{dsfont}

\usepackage[]{jabbrv}

\usepackage[numbers,sort&compress]{natbib}

\graphicspath{{./}{./figures/}}

\newcommand{\vect}[1]{\vec{#1}} 
\newcommand{\tensor}[1]{  {\bm{#1}} } 

\newcommand{\vm}[1]{\vect{#1}} 

\newcommand{\divergence}[1]{ {\rm div} \left[ \, {#1} \, \right] }
\newcommand{\gradient}[1]{ {\rm \nabla} \left[ \, {#1} \, \right] }

\newcommand{\curl}[1]{ {\rm{curl}} \left[ \, {#1} \, \right] }

\newcommand{\gradientone}[1]{ \frac{ \partial {#1} }{\partial x} }

\newcommand{\myvarepsilon}{\mbox{$\varepsilon \mskip-6mu  | $}}

\newcommand{\Temperature}{T}

\newcommand{\diffusivity}{\mbox{${\rm D} \mskip-8mu  | \,$}}
\newcommand{\ionmobility}{\mbox{${\rm u} \mskip-8mu  | \,$}}
\newcommand{\elchempot}{ \overline{\mu}}

\newcommand{\ym}{ {y}}
\newcommand{\ymh}{ {\hat{y}}}

\newcommand{\redcom}[1]{{\color{red} {#1}}}

\title{A two-mechanism and multiscale compatible approach for solid state electrolytes of (Li-ion) batteries. }

\author[1]{L. Cabras}
\author[2,3]{D. Danilov}
\author[4]{W. Subber}
\author[5]{V. Oancea}
\author[1]{A. Salvadori}

\affil[1]{Dipartimento di Ingegneria Meccanica e Industriale, Universit\`a di Brescia, Italy}
\affil[2]{Eindhoven University of Technology, P.O. Box 513, 5600 MB Eindhoven,  The Netherlands}
\affil[3]{Forschungszentrum J{\"u}lich, Fundamental Electrochemistry (IEK-9), D-52425 Jülich,  Germany}
\affil[4]{GE Research, Niskayuna NY 12309, United States}
\affil[5]{Dassault Systemes Simulia Corp, United States}

\begin{document}

\maketitle

\begin{abstract}
All solid state batteries are claimed to be the next-generation battery system, in view of their safety accompanied by high energy densities.
A new advanced, multiscale compatible, and fully three dimensional model for solid electrolytes is presented in this note.
The response of the electrolyte is profoundly studied theoretically and numerically, analyzing the equilibrium and steady state behaviors,
the limiting factors, as well as the most relevant constitutive parameters according to the sensitivity analysis of the model.
\end{abstract}


\section{Introduction}
\label{sec:intro}

All solid state batteries (SSBs) are claimed to be the next-generation battery system, since they combine superior thermal and electrochemical stability and avoid hazardous liquid electrolyte leakage \cite{SCHNELL2018160,Zheng2018}. 
As pointed out extensively in \cite{bistriM2020}, SSBs still present a number of chemical and stability issues. In conjunction with experimental campaigns, thermodynamically grounded models and simulations can provide insights into battery operation, limiting factors, and degradation. The ability to understand the physics behind irreversible processes, which ultimately leads to capacity fade, is related to the availability of 
accurate models for solid electrolytes and electrodes. These models shall 
incorporate several phenomena, which are interconnected at different scale during batteries operations \cite{LiMonroeARCBM2020}.

In a companion paper \cite{CabrasEtAl2021b} we carefully reviewed three notable papers on SSBs, displaying the evolution of cornerstone ideas on the solid electrolyte.
At standard conditions, some of the Li ions are thermally excited within the solid electrolyte. Chemical ionization reactions occur, leaving behind uncompensated negative charges, associated with a vacancy in the matrix at the place formerly occupied by lithium.
In most cases, see e.g. \cite{Fabre2012} and the references therein, the ionic transfer is described by a {\em{single ion conduction model}}. Since the negative vacancies in the lattice are modeled as firmly held, they cannot flow and 
the resulting concentration of Li ions across the solid electrolyte is uniform and known a priori in view of the electroneutrality \cite{SalvadoriEtAlIJSS2015}. Because no concentration gradient drives the ionic motion, those models reproduce essentially Ohm's law. 

The single ion conduction models have been enriched by the description of the {\em{interfaces mechanisms}}. Intermediate electrode/electrolyte layers have been modeled as interfaces between electrodes and the solid electrolyte, in terms of potential jumps as for plate capacitors \cite{BonnefontEtAlJEC2001}. Stemming from rigorous thermodynamic setting, conditions of non Butler-Volmer type arise \cite{LandstorferEtAlPCCP2011}.

More recently, single ion conduction models have been displaced by {\em{two-mechanism models}}, which describe with more realism the ionic motility in the solid electrolyte. One-dimensional mathematical models for Lithium phosphorus oxynitride ($\rm LiPON$ henceforth) have been proposed in \cite{Danilov2016A,Danilov2016B,RaijmakersEtAlEA2020}, whereas a novel two-mechanisms model was illustrated in \cite{CabrasEtAl2021b} and validated against experimental evidences published in \cite{Danilovetal2011}. Although simulations reproduce well
the behavior of a cell, deep analyses of the model have not been elaborated in \cite{CabrasEtAl2021b}. The present note closes such a gap by carrying out a detailed, profound theoretical and numerical investigation of the response of the electrolyte, analyzing the equilibrium and steady state behaviors,
the limiting factors of the model, as well as the most relevant constitutive parameters according to the sensitivity analysis (SA).

In  \cite{RaijmakersEtAlEA2020} both interstitial lithium and negative vacancies were allowed to flow, thus creating a concentration gradient at steady state that resembles the liquid electrolyte distributions found for instance in \cite{SalvadoriEtAlJPS2015,SalvadoriEtAlJPS2015b}.  Depicting vacancies with the same conceptual framework used for negative ions in liquid electrolytes, i.e. as able to move in the solid matter with an entropic brownian motion together with migration within an electric field, does not appear to be physically sound. 
In our formulation, we model explicitly the dynamic filling of vacancies by neighboring positions, a motion of positive ions which in turn creates new vacancies. To this aim, we claim that after the ionization reactions occur, some ions hop and fill neighboring vacancies, whereas the remaining positive ions move in a meta-stable interstitial state. In this way, positive ions are the only moving species and the concentration of negatively charged vacancies results from the solution of the governing equations. Such a set of partial differential equations has been detailed in section \ref{sec:locbalancelaws} and is made of mass balance equations, chemical kinetics laws, balance of momentum, and Ampere's law. Those continuity equations shall be supplied with constitutive laws, which arise
from a rigorous thermodynamic analysis formulated in section \ref{sec:constitutivetheory}. 
Numerical simulations via the finite element method (FEM) permit to recover the steady state response of the system as well as the transient path of the unknown fields when initial conditions are far from equilibrium (see section \ref{subsec:transient}), a typical situation in real batteries. Since model validation against experimental evidence has been carried out in \cite{CabrasEtAl2021b}, we did not indulge here on this matter.

Governing equations can be solved rather straightforwardly at steady state as well as at equilibrium: the closed form solutions highlight the role of material parameters, some of which can be measured only with major uncertainties. 
SA helps in identifying model parameters that contribute the most to the prediction, and thus identifying the accuracy required in measuring these parameters \cite{saltelli2004sensitivity}. 
The SA carried out in section \ref{sec:sensitivity} allows to figure out the effect of the variability of model parameters on the variability of its prediction. The SA suggests that the fraction of Li that resides in equilibrium in the mobile state is the most sensitive parameter.

%

\section{Electrochemical modeling of the solid electrolyte}
\label{sec:locbalancelaws}

The model proposed in \cite{RaijmakersEtAlEA2020} inspired this novel two-mechanism study, which is grounded in the thermo-mechanics of continua. It advances \cite{RaijmakersEtAlEA2020} in modeling the process of vacancies replenishment and in making it multi-scale-compatible, which appears to be relevant for composite cathodes \cite{Fathiannasab_2020,FATHIANNASAB2021229028,BielefeldEtAlJPC2019}. The version of the model detailed in what follows is only a restriction of a broader multi-physics formulation, which includes mechanical and thermal interactions according to \cite{SalvadoriEtAlJMPS2018}. However, since the present note concerns the electrochemical performance, for the sake of conciseness we neglect here those interactions, which will be elaborated in further publications. Henceforth we will assume both thermal and mechanical equilibrium, with relevant fields fixed during electrolyte operation.

%
\begin{figure}[!htb]
\centering
\begin{subfigure} {0.495\textwidth}
  \includegraphics[width=\textwidth]{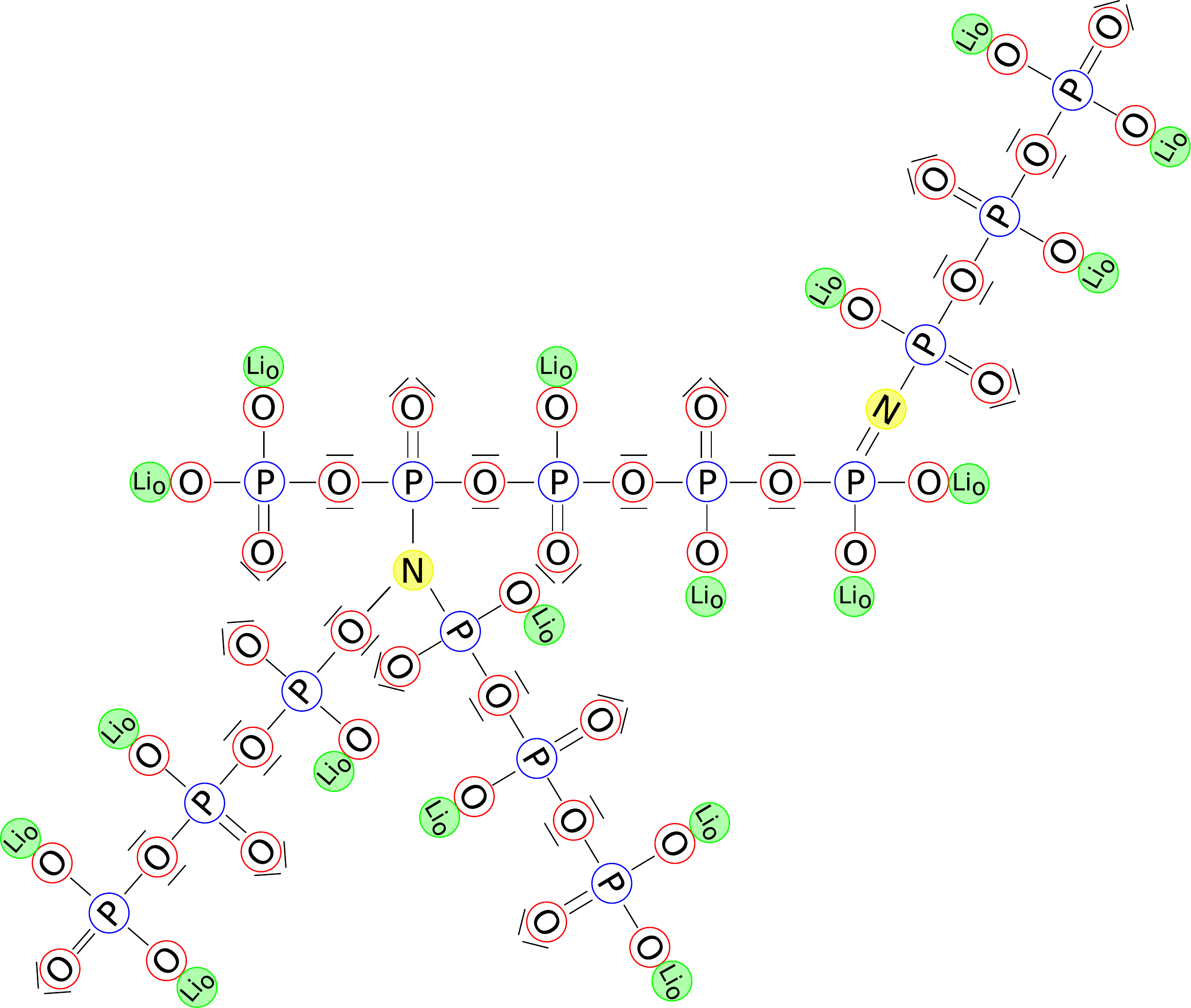}
\caption{  }
\end{subfigure}
\begin{subfigure} {0.495\textwidth}
  \includegraphics[width=\textwidth]{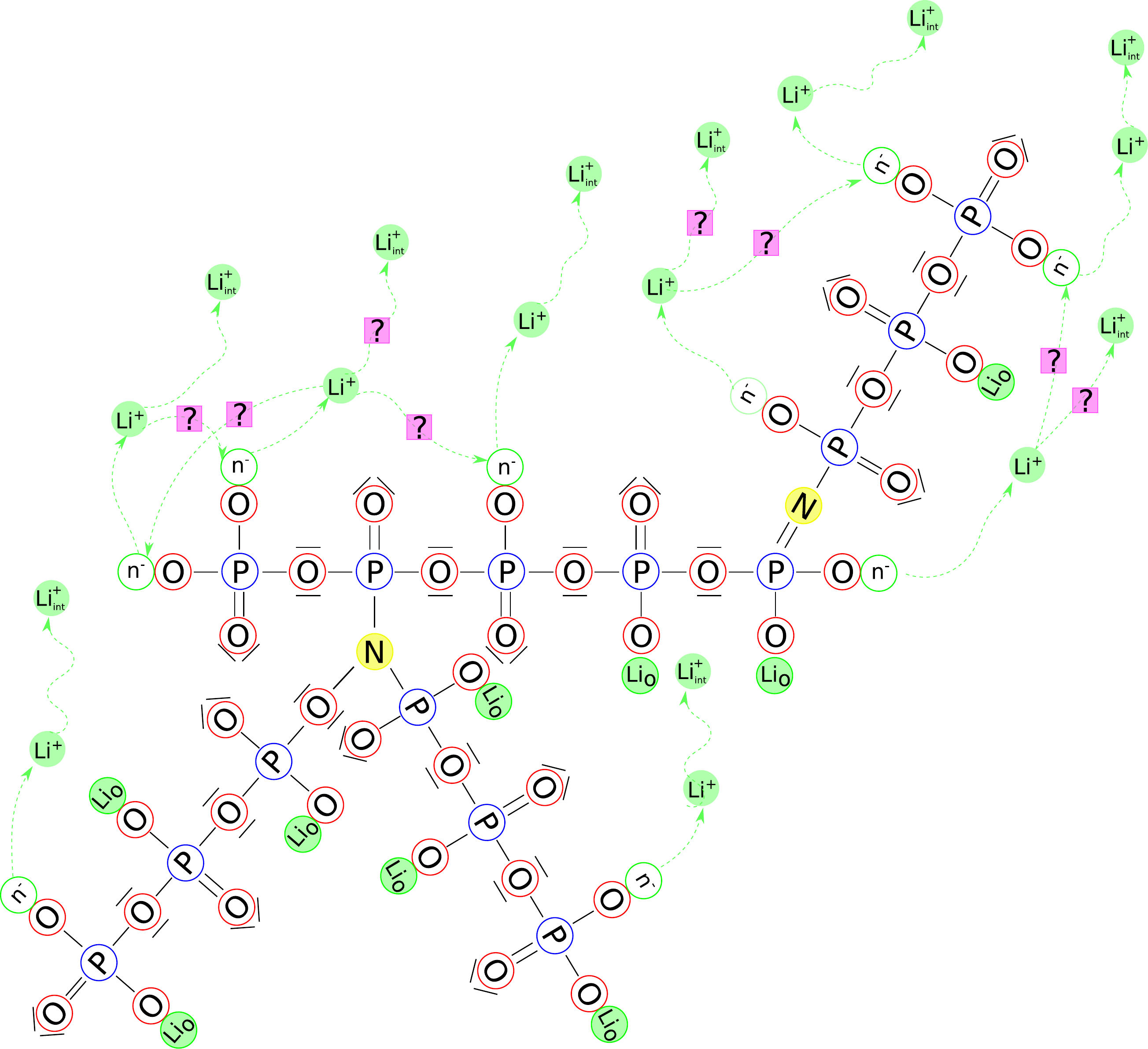}
\caption{  }
\end{subfigure}

\caption{\em LiPON matrix with triply- and doubly coordinated nitrogen (a). Movements of charged particles towards the interstitial space and by means of particle hopping, representing the main ionic conductivity mechanisms in LiPON (b).}
\label{fig:ReticoloDanilov}
\end{figure}
%
%

\subsection{ Chemical kinetics}
\label{subsec:chem_kin}

The amorphous structure of the LiPON electrolyte is schematically shown in Fig. \ref{fig:ReticoloDanilov}. It highlights two types of nitrogen bonds, either triply- or doubly coordinated.
 $\rm Li_0$ denotes the (ionic) lithium bound to the non-bridging oxygen atoms, $\rm Li^+$ is a lithium ion and $\rm n^-$ is the uncompensated negative charge associated with a vacancy formed in the LiPON matrix at the place where $\rm Li^+$ was originally bound. The maximal concentration of host-sites, denoted with $c_0$, is established by the stoichiometric composition of the electrolyte material. It is reached in the ideal case of absolute zero temperature, when all available host sites are fully filled with lithium ions and the ionic conductivity vanishes because all ions are immobile, see Fig. \ref{fig:ReticoloDanilov}a. In standard conditions, see Fig. \ref{fig:ReticoloDanilov}b, some of the Li-ions are thermally excited and the chemical ionization reaction
\begin{equation}
{\rm Li_0} \underset{k_b^{\rm ion}}{\overset{k_f^{\rm ion}}{\rightleftarrows}} {\rm Li^+ + n^-}
\label{eq:IonizationReaction1}
\end{equation}
occurs, $k_f^{\rm ion}$ and $k_b^{\rm ion}$ being the forward and backward rate constants for the ionization (or recombination) reaction, respectively. Their ratio is the equilibrium constant of reaction \eqref{eq:IonizationReaction1}
\begin{equation}
\label{eq:Keqion}
K_{\rm eq}^{\rm ion} = \frac{ k_f^{\rm ion} }{ k_b^{\rm ion} }
\; .
\end{equation}

%

The ionization reaction \eqref{eq:IonizationReaction1} leaves behind uncompensated negative charges associated with a vacancy in the LiPON matrix at the place formerly occupied by lithium. In \cite{RaijmakersEtAlEA2020} those vacancies were modeled with the same conceptual formalism used for negative ions in liquid electrolytes, i.e. as able to move in the solid matter driven by an entropic Brownian motion together with migration within an electric field. 
Here, we attempt to explicitly account for the dynamic filling of vacancies by neighboring positions, where new vacancies are created. To this aim, we claim that after the chemical ionization reaction \eqref{eq:IonizationReaction1} occurs, some ions, denoted henceforth with ${\rm Li}^+_{\rm hop}$, hop and fill neighboring vacancies, whereas the remaining $\rm Li^+$ ions move in a meta-stable interstitial state. This dynamic behavior is described by a further reaction, that converts part of the full amount of ions made available by reaction \eqref{eq:IonizationReaction1} into hopping lithium with the ability to fill vacancies, leaving to the remaining ions the interstitial motion responsibility :
\begin{equation}
{\rm Li^+} \underset{k_b^{\rm hop}}{\overset{k_f^{\rm hop}}{\rightleftarrows}} {\rm Li}^+_{\rm hop}
\; .
\label{eq:IonizationReaction2}
\end{equation}
$k_f^{\rm hop}$ and $k_b^{\rm hop}$ are the rate constants for reaction \eqref{eq:IonizationReaction2}.  Their ratio is the equilibrium constant of reaction \eqref{eq:IonizationReaction2}
\begin{equation}
\label{eq:Keqhop}
K_{\rm eq}^{\rm hop} = \frac{ k_f^{\rm hop} }{ k_b^{\rm hop} }
\; .
\end{equation}

Therefore, reaction \eqref{eq:IonizationReaction1} makes lithium ions capable of unbinding from the non-bridging oxygen atoms and move within the complex amorphous LiPON structure, either by filling neighboring vacancies or by flowing interstitially. The proportion of ions in these two mechanisms is governed by reaction \eqref{eq:IonizationReaction2}. When $k_f^{\rm hop} = 0$, no hopping mechanism is accounted for.
As $k_f^{\rm hop} $ increases, since more interstitial lithium is depleted going for hopping, more vacancies are formed in the ionization reaction  \eqref{eq:IonizationReaction1}  thus favoring the hopping mechanism.

\bigskip
Denote with $c_{{\rm Li}^+}$ and $c_{{\rm Li}^+_{\rm hop}}$ the (molar, i.e. the number of moles per unit volume) concentration of mobile Li ions, with $c_{\rm Li_0}$ the concentration of immobile lithium, with $c_{\rm n^-}$ the concentration of uncompensated negative charges. Concentrations $c_{\alpha}$ express the {\em{molarity}} (i.e. the number of moles per unit volume) of a generic species $\alpha$; $\vect{h}_\alpha$ is the mass flux in terms of moles, i.e. the number of moles of species $\alpha$ measured per unit area per unit time.  Scalar and vector fields are defined in space $\vect{x} \in V$ and time $0 \le t \le t_f$, i.e. $c_{\alpha}=c_{\alpha}(\vect{x},t)$. Functional dependence, however, is specified when necessary only to enhance readability.

\bigskip
For ideal systems, in which  chemical potentials have entropy and {energy} contributions only, the chemical kinetics of reactions \eqref{eq:IonizationReaction1} and \eqref{eq:IonizationReaction2} are modeled via the law of mass action \cite{deGrootBook}:
\begin{equation}
        w = k_f^{\rm ion} \; \frac{ \theta_{{\rm Li_0}} }{ 1-\theta_{{\rm Li_0}} } -  k_b^{\rm ion} \; \frac{ \theta_{{\rm Li}^+} }{ 1-\theta_{{\rm Li}^+} } \; \frac{ \theta_{\rm n^-} }{ 1-\theta_{\rm n^-} }
      \; ,
 \label{eq:idealmassactionlaw}
\end{equation}
where $\theta_\alpha$ is the dimensionless ratio 
$$
 \theta_\alpha = \frac{ c_\alpha }{ c^{sat}_\alpha }
$$
and $c^{sat}_\alpha$ is the saturation limit of the generic species $\alpha$. In diluted conditions, i.e. $\theta_\alpha \ll 1$, eq. \eqref{eq:idealmassactionlaw} writes
\begin{subequations}
\begin{equation}
	 w(\vect{x},t) = k_f^{\rm ion} \; c_{\rm Li_0}(\vect{x},t) - k_b^{\rm ion} \; c_{{\rm Li}^+}(\vect{x},t) \; c_{\rm n^-}(\vect{x},t)
      \; ,
 \label{eq:mass_action_ion}
\end{equation}
with a small abuse of notation on the kinetic constants. The reaction rate of the interstitial-hopping transformation is
\begin{equation}
	y(\vect{x},t) =k_f^{\rm hop} \; c_{{\rm Li}^+}(\vect{x},t)  -  k_b^{\rm hop} \; c_{{\rm Li}^+_{\rm hop}}(\vect{x},t)
	\; .
\label{eq:mass_action_hop}
\end{equation}

As observed in \cite{SalvadoriEtAlJMPS2018} for 
the elastic and swelling contributions, electric potential may affect the kinetics of reaction and hence the law of mass action. Consistently with \cite{SalvadoriEtAlJMPS2018} and the Arrhenius formalism,   it is proposed here that factors $k_f$ and $k_b$ in eqs. \eqref{eq:mass_action_ion}, \eqref{eq:mass_action_hop} are function of the electric potential $\phi$ in the following way:
\begin{align}
  & k_f^{\rm ion} = k_{f_0}^{\rm ion} \; e^{   \frac{ \zeta F \phi}{RT} }   \; ,
  & k_f^{\rm hop}=k_{f_0}^{\rm hop} \; e^{   \frac{ \zeta F \phi}{RT}  }   \; ,
 \\ 
 & k_b^{\rm ion}=k_{b_0}^{\rm ion} \; e^{  \frac{ \zeta F \phi}{RT}  }   \; ,
 & k_b^{\rm hop}=k_{b_0}^{\rm hop} \; e^{\frac{ \zeta F \phi}{RT} }   \; ,
\end{align}
 \label{eq:kasfunctionofphi}
\end{subequations}
with $R = 8.31446	\, \rm J \, K^{-1} mol^{-1}$ the gas constant, $T$ temperature, $F= 96485.338 \, \; {\rm{C} \, mol}^{-1}$ Faraday's constant, $k_{f_0}$ and $k_{b_0}$ positive constants, $\zeta$ an amplification factor. When $\zeta=0$, the influence of the electric potential vanishes.
This new formulation is consistent with the usual mass action law, which is recovered when the potential equals the reference potential, here taken as zero. Note that the equilibrium constants $K_{\rm eq}$ in eqs.  \eqref{eq:Keqion} or \eqref{eq:Keqhop} remain independent upon the electric potential, which thus influences the velocity of the two reactions but not their equilibrium state.

\subsection{Mass balance}
\label{subsec:massbalance}

In this formalism, positive ions are the only moving species, whereby the concentration of vacancies is the outcome of the motion process; uncompensated negative charges do not possess any intrinsic motility and some of them are filled by ${\rm Li}^+_{\rm hop}$ ions. In this sense, there is no direct flow $\vect{h}_{\rm n^-}(x,t)$ of negative charges and the local concentration of vacancies is altered merely by the chemical ionization reaction  eq. \eqref{eq:IonizationReaction2}.

In view of equation \eqref{eq:IonizationReaction1}, each lithium ion that leaves the host site creates a negatively charged uncompensated vacancy. Therefore, since the amorphous structure is not suppose to reorder itself, the concentration of the vacancies plus the concentration of immobile lithium shall remain constant in time and equal the maximal concentration of host-sites
\begin{equation}
c_{\rm Li_0} + c_{\rm n^-} = c_0
\; .
\label{eq:host_sites_conservation}
\end{equation}
%

%
%

%
%
%

The mass balance equations characterize the chemo-diffusive migration transport of species within the solid electrolyte. Continuity equations are stated in a general, three-dimensional framework, although applications in this note will be merely one-dimensional (see  Section \ref{sec:numerical simulation}). 
%
For the immobile lithium $\rm Li_0$ and for uncompensated negative charges $\rm n^-$, the mass balance equations are made distinctive by the absence of fluxes
\begin{subequations}
\begin{align}
 \frac{ \partial c_{\rm Li_0}}{ \partial t}=-w,\\
 \frac{ \partial c_{\rm n^-}}{ \partial t}=w,
\end{align}
\label{eq:MassEle_model1}
\end{subequations}
where the overall rate of the charge carrier generation under general (dynamic) conditions is given by the mass action law \eqref{eq:mass_action_ion}.
For the interstitial and hopping lithium ions the mass balance equations read
\begin{subequations}
\begin{align}
 \frac{\partial c_{{\rm Li}^+}}{ \partial t}+\divergence{\vm{h}_{{\rm Li}^+}}=w-y,\\
 \frac{ \partial c_{{\rm Li}^+_{\rm hop}}}{ \partial t}+\divergence{ \vm{h}_{{\rm Li}^+_{\rm hop}}}=y,
\end{align}
\label{eq:MassEle_model2}
\end{subequations}
where the reaction rate of the interstitial-hopping transformation is depicted by eq. \eqref{eq:mass_action_hop}.

\subsection{Charge balance}
\label{subsec:FaradayLaw}

Charges in the solution are due to negatively charged uncompensated vacancies as well as to the transport of interstitial and hopping positive ions:
\begin{subequations}
\begin{equation}
\label{eq:chargeandmass}
  {\zeta} = F \, \left(  c_{{\rm Li}^+} + c_{{\rm Li}^+_{\rm hop}} -  c_{\rm n^-} \right)
  \; .
\end{equation}
%
The flux of mass in balance (\ref{eq:MassEle_model2}) of each species contributes to a current density $\vect{i}$
\begin{equation}
  \vect{i} = F \, \left( \vm{h}_{{\rm Li}^+} + \vm{h}_{{\rm Li}^+_{\rm hop}} \right)
  \; .
\label{eq:FaradaysLaw}
\end{equation}
\end{subequations}

\subsection{Maxwell's equations for electro-quasi-statics}
\label{subsec:GaussLaw}

The model identified so far involves 4 different species,  whose concentrations are $c_{\rm Li_0}$, $c_{\rm n^-}$, $c_{{\rm Li}^+}$, and $c_{{\rm Li}^+_{\rm hop}}$. The set of 4 mass balance equations, \eqref{eq:MassEle_model1} and \eqref{eq:MassEle_model2}, contains 5 unknowns, i.e. the 4 mass concentrations plus the electric potential, which is constitutively related to the mass fluxes. An additional equation is required and a common selection in battery modeling is the {\em{electroneutrality}} condition (see among others \cite{NewmanBook2004}, page 286), which for the model at hand reads
\begin{equation}
c_{{\rm Li}^+}(\vect{x},t) + c_{{\rm Li}^+_{\rm hop}}(\vect{x},t)  = c_{\rm n^-}(\vect{x},t)
\; .
\label{eq:electroneutrality}
\end{equation}
In several studies, originated by Newman \cite{NewmanBook2004} and collectively gathered in the terminology ``porous electrode theory'', condition (\ref{eq:electroneutrality}) is used {{in place of Maxwell's law}} - see among others \cite{DOYLEETALJES1993, Garcia2005, WangSastry2007, GolmonEtAlCS2009, Christensen2010, RenganathanEtAl2010, GolmonEtAlIJNME2012,DanilovNottenEA2008, Danilovetal2011,  ZadinEtAl2012}. Taking advantage of eq. \eqref{eq:electroneutrality}, the electric
field is not constrained in any way to satisfy Maxwell's equations. Remarkably, electroneutrality does violate\footnote{This is immediately accomplished in 1D, where in view of electroneutrality Gauss law reads $\gradientone{D}=0$, thus leading to a constant electric field. } Maxwell's equations (see for instance \cite{RademakerEtAlJES2014}).

As discussed in \cite{SalvadoriEtAlIJSS2015,SalvadoriEtAlJPS2015}, in multiscale approaches electroneutrality assumption (\ref{eq:electroneutrality}) cannot be used, since it does not allow to ensure energy conservation in the scales transitions\footnote{In fact, in a rigorous multi scale mathematical formulation - see \cite{Suquet1985, GEERSJCAM2010} - the micro to macro scale transition requires that the same power is expended at the two scales, thus assuring that energy is neither artificially generated nor artificially dissipated across the scales. If electroneutrality is used in place of Maxwell's equations, recovering the energy description of the electromagnetic interactions is hardly possible.}. This is a major obstacle to the development of predictive theories for the battery response with multi-scale models \cite{FrancoRCS2013,FrancoBook2016,FrancoEtAlCR2019,Dargaville2010,LatzZauschBJN2015,LiMonroeARCBM2020,Pannala2015,Wieser2015,zielkeEtAlAEM2015}.

In the batteries modeling literature \cite{NewmanBook2004, BardFaulknerBook, HugginsBook2010}, it is generally assumed that the electromagnetic fields and their interactions
are {\em{static}}. This assumption implies vanishing interference effects between the electric and magnetic phenomena. As a consequence, the set of Maxwell's equations are replaced by their electrostatic counterparts, as for the steady current case \cite{LandauLifshitz1984}.
In the present paper, equation (\ref{eq:electroneutrality}) is not used as a fundamental law. Instead, electromagnetics is explicitly taken into account via the {\em{electro-quasi-static}} formulation \cite{LarssonAJP2007} of Maxwell's equations\footnote{As insightfully noticed in \cite{LarssonAJP2007}, electrostatics is a particular case of the general Maxwell's equations but electro-quasi-statics is not, it is {\em{an approximation}}. Such an approximation is acceptable under some conditions, described in \cite{HausMelcherBook1989}.}, following the same path of reasoning of \cite{SalvadoriEtAlIJSS2015}. 
By this approach, the time-dependent hyperbolic Maxwell's equations are replaced by parabolic equations that can be solved in a more simple way.

\bigskip
Gauss's laws relate the electric displacement and magnetic fields ($\vect{D}$ and $\vect{B}$ respectively) emanating from the distribution \eqref{eq:chargeandmass} of electric charge $\zeta$
\begin{eqnarray}
\label{eq:GaussLaw}
    \divergence{\vect{D}} =  \zeta  \; ,
\qquad
    \divergence{\vect{B}} = 0
    \; .
\end{eqnarray}
In the light of the simplification made,  the time derivative of the magnetic field is negligible within Maxwell-Faraday's law of induction, hence the electric field is irrotational and derives from an {\em electrostatic potential} $\phi$:
\begin{equation}
\label{eq:ElectrostaticPotential}
    \vect{E} = - \gradient{\phi}
    \; .
\end{equation}
Finally, Amp{\`e}re's law (with Maxwell's correction)
\begin{equation}
\label{eq:AmpereLaw}
      \frac{\partial \vect{D}}{\partial \,t} + \vect{i} = \curl{\vect{H}}
\end{equation}
%
relates the electrical current \eqref{eq:FaradaysLaw} and the time variation of the electric displacement field to the {\em{magnetizing field}} $\vect{H}$. The impingement of the latter in Amp{\`e}re's law cannot be disregarded in the simplified framework of electro-quasi-statics.  Nonetheless, a differential form can be straightforwardly obtained from Amp{\`e}re's law (\ref{eq:AmpereLaw}), after application of the divergence operator:
\begin{equation}
\label{eq:IncrementalGaussLaw}
     \divergence{ \frac{\partial \vect{D}}{\partial \,t} + \vect{i} } = {0}
           \; .
\end{equation}
This form will be used in the sequel of the paper.
%
%

\subsection{Weak form and boundary conditions}
\label{sec:MicroScaleWeakForm}

A weak form that entails a proper energy meaning can be given as in  \cite{SalvadoriEtAlIJSS2015} multiplying the strong form of the mass balance equations by a suitable set of chemical potentials test functions (${\hat \mu}_{{\rm Li_0}}$, ${\hat \mu}_{{\rm n}^-}$, ${\hat \mu}_{{\rm Li}^+}$, ${\hat \mu}_{{\rm Li}^+_{\rm hop}} $) and performing integration by parts, exploiting Green's formula with the aim of reducing the order of differentiation.
The weak form of the mass balance equations (\ref{eq:MassEle_model1},\ref{eq:MassEle_model2}) for all the species read:
\begin{subequations}
\begin{align}
&
\int_{V} {\hat \mu}_{\rm Li_0} \; { \frac{ \partial c_{\rm Li_0} }{ \partial t} \,}{\rm d}V = - \int_{V} {\hat \mu}_{\rm Li_0} \; w \; {\rm d}V
\; ,
\\
&
\int_{V} {\hat \mu}_{\rm n^-} \; { \frac{ \partial c_{\rm n^-} }{ \partial t} \,}{\rm d}V = \int_{V} {\hat \mu}_{\rm n^-} \; w \; {\rm d}V
\; ,
\\
&
\int_{V}
{\hat \mu}_{{\rm Li}^+} \;   \frac{ \partial c_{{\rm Li}^+} }{ \partial t} \,
-
\gradient{ {\hat \mu}_{{\rm Li}^+} } \cdot  \vm{h}_{{\rm Li}^+} \, {\rm d}V
+
\int_{\partial V}{ {\hat \mu}_{{\rm Li}^+} \;  \vm{h}_{{\rm Li}^+} } \cdot {\vect{n}} \; {\rm d}\Gamma
=
\int_{V} {\hat \mu}_{{\rm Li}^+} \;  ( w-y ) \; {\rm d}V
\; ,
\\
&
\int_{V}
{\hat \mu}_{{\rm Li}^+_{\rm hop}} \;   \frac{ \partial c_{{\rm Li}^+_{\rm hop}} }{ \partial t} \,
-
\gradient{ {\hat \mu}_{{\rm Li}^+_{\rm hop}} } \cdot  \vm{h}_{{\rm Li}^+_{\rm hop}} \, {\rm d}V
+
\int_{\partial V}{ {\hat \mu}_{{\rm Li}^+_{\rm hop}} \;  \vm{h}_{{\rm Li}^+_{\rm hop}} } \cdot {\vect{n}} \; {\rm d}\Gamma
=
\int_{V} {\hat \mu}_{{\rm Li}^+_{\rm hop}} \;  y \; {\rm d}V
\; .
\end{align}
\label{eq:weakmicromassbalance}
\end{subequations}
Two terms at the left-hand side of (\ref{eq:weakmicromassbalance}c,d) are defined at the boundary $\partial V$; $\vect{n}$ is the outward normal to the surface of the electrodes.
The electrolyte boundaries, forming the interfaces with electrodes, are of major interest in energy storage systems. A large amount of research has been devoted to modeling the
electrical double layer at solid-state electrochemical interfaces \cite{SwiftEtAlNCS2021}.
%
%
As this work is restricted to the electrolyte only, electrode kinetics is not detailed and we rather refer to a companion paper \cite{CabrasEtAl2021b}. 
The weak form however clearly points out the need of splitting the lithium flux at the boundary into two terms,
\begin{subequations}
\begin{align}
  \vm{h}_{{\rm Li}^+} \cdot \vect{n} = - h^{BV}_{{\rm Li}^+}  \qquad \vect{x} \in \partial^N V
  \; ,
    \\
  \vm{h}_{{\rm Li}^+_{\rm hop}} \cdot \vect{n} = - h^{BV}_{{\rm Li}^+_{\rm hop}}  \qquad \vect{x} \in \partial^N V
  \; ,
\end{align}
\label{eq:MassFluxBC}
\end{subequations}
where the mass fluxes at the boundary, termed $h^{BV}_{{\rm Li}^+}$ and $h^{BV}_{{\rm Li}^+_{\rm hop}}$, must descend from a proper interface equation, generally of Butler-Volmer type. 
In this note the amounts of $h^{BV}_{{\rm Li}^+}$ and $h^{BV}_{{\rm Li}^+_{\rm hop}}$ will be taken as given terms.

\bigskip
With a similar path of reasoning and accounting for eq. \eqref{eq:FaradaysLaw}, the weak form of Amp{\`e}re's law (\ref{eq:IncrementalGaussLaw}) reads
\begin{align}
\int_{V}
-
\gradient{ {\hat \phi}} \cdot
\left\{ \frac{\partial {\vm{D}}}{\partial t}  + F  \left (\vm{h}_{{\rm Li}^+}+\vm{h}_{{\rm Li}^+_{\rm hop}} \right) \right\}
{\rm d}V
+
\int_{\partial V}
{\hat \phi}
\;
\left\{  \frac{\partial {\vm{D}}}{\partial t}  +  F  \left (\vm{h}_{{\rm Li}^+}+\vm{h}_{{\rm Li}^+_{\rm hop}}\right )\right\}
\cdot
{\vect{n}} \, {\rm d}\Gamma = {0}
\; .
\label{eq:weakmicromassbalance2}
\end{align}

Boundary conditions for the electric potential emanate from Amp{\`e}re's law (\ref{eq:AmpereLaw}), accounting for constraints \eqref{eq:MassFluxBC}.
\begin{equation}
\label{eq:magncoupl}
 \left\{   \frac {\partial \vm{D}}{\partial t} + F \,  \left( \vm{h}_{{\rm Li}^+} + \vm{h}_{{\rm Li}^+_{\rm hop}} \right)  \;   \right\}  \cdot \vect{n} = \curl{ \vm{H}} \cdot \vect{n}
 \qquad \vect{x} \in \partial V
 \; .
\end{equation}
In the modeling a full battery cell, it can be assumed that the curl of the magnetizing field is continuous across all interfaces when projected in the normal direction. Such a continuity condition cannot be rephrased for the case where only the electrolyte is modeled.
It will be assumed henceforth that $\vm{B}$ {\em along the boundary} can be estimated from the ``steady current" theory (see \cite{LandauLifshitz1984}, chapter 3).  Amp{\`e}re's law without Maxwell's correction describes the magnetic field generated by a steady current
\begin{equation}
\label{eq:curlbc}
   \curl{ \vm{H} } \cdot \vect{n} = - F \, \left (h^{BV}_{{\rm Li}^+}+h^{BV}_{{\rm Li}^+_{\rm hop}}\right )  \qquad \vect{x} \in \partial^N V
   \; .
\end{equation}
In view of (\ref{eq:MassFluxBC}) and (\ref{eq:magncoupl}), boundary conditions for the electric potential read:
\begin{align}
     \frac {\partial \vm{D}}{\partial t}   \cdot \vect{n} = 0 \qquad \vect{x} \in \partial^N V
     \; .
\label{eq:ElPotBC}
\end{align}
Note however that this condition {\em{is not imposed}} in full cells. Finally, in order to make the problem solvable, Dirichlet boundary conditions (usually homogeneous) for the potential need to be added.

\bigskip
In conclusion, the weak form of the {\em{balance}} equations can be written in terms of the potentials in time interval $\left[ 0, t_f \right]$ as
\begin{eqnarray}
\mbox{ Find } z \in \mathcal{V}^{[ 0,t_f ]}  \mbox{ such that }
\hspace{1cm}
 \frac{\rm d}{ {\rm d}t} b \left( \ymh,  {z}(t) \right) + a ( \ymh,  {z}(t) ) + c ( \ymh,  {z}(t) )= f(\ymh)
\hspace{1cm}
 \forall \ymh \in \mathcal{V}
 \label{eq:microweakform}
\end{eqnarray}
where
\begin{subequations}
\begin{align}
b \left( \ymh, {z} \right) &=
   \int_{V} \;   {\hat { \mu}}_{{\rm Li_0}} \, c_{{\rm Li_0}} +{\hat { \mu}}_{{\rm n}^{\!-}} \, c_{{\rm n}^-}+{\hat { \mu}}_{{\rm Li}^+} \, c_{{\rm Li}^+}+{\hat { \mu}}_{{\rm Li}^+_{\rm hop}} \, c_{{\rm Li}^+_{\rm hop}} \,	{\rm d}V
    - \int_{V} \;   \gradient{  {\hat \phi} } \cdot  \vm{D}  \, {\rm d}V
 \\
a \left( \ymh, {z}(t) \right)  &=
    -\int_{V} \;\gradient{  {\hat { \mu}}_{{\rm Li}^+} } \cdot \vm{h}_{{\rm Li}^+} +\gradient{  {\hat { \mu}}_{{\rm Li}^+_{\rm hop}} } \cdot \vm{h}_{{\rm Li}^+_{\rm hop}}{\rm d}V +\int_{V} \;   \gradient{  {\hat \phi} } \cdot  F \,  \left( \vm{h}_{{\rm Li}^+} + \vm{h}_{{\rm Li}^+_{\rm hop}} \right)   \,	{\rm d}V
\\
c \left( \ymh, {z}(t) \right)  &=
    \int_{V} \;{\hat { \mu}}_{{\rm Li_0}}\cdot w\,{\rm d}V-\int_{V} \;{\hat { \mu}}_{{\rm n}^-}\cdot w\,{\rm d}V-\int_{V} \;{\hat { \mu}}_{{\rm Li}^+}\cdot (w-y)\,{\rm d}V -\int_{V} \;{\hat { \mu}}_{{\rm Li}^+_{\rm hop}}\cdot y\,{\rm d}V
\\
f \left( \ymh \right) &=
 - \int_{\partial^N V } \; {\hat { \mu}}_{{\rm Li}^+} {h}^{BV}_{{\rm Li}^+}+{\hat { \mu}}_{{\rm Li}^+_{\rm hop}} {h}^{BV}_{{\rm Li}^+_{\rm hop}}-F \hat{ \phi}\left ({h}^{BV}_{{\rm Li}^+}+{h}^{BV}_{{\rm Li}^+_{\rm hop}}\right) \; {\rm d} \Gamma	
\end{align}
\label{eq:weakmicromassbalance4}
\end{subequations}
with ${ z}= \{\, c_{{\rm Li_0}} , \, c_{{\rm n}^-} ,  \, c_{{\rm Li}^+} , \, c_{{\rm Li}^+_{\rm hop}},\phi \}$,  $\ym = \{{ \mu}_{{\rm Li_0}}, { \mu}_{{\rm n}^-},{ \mu}_{{\rm Li}^+},{ \mu}_{{\rm Li}^+_{\rm hop}},\phi \}$.
Columns ${ z}$ and $\ym$ collect the time-dependent unknown fields. Column $\ymh$ collects the steady-state test functions that correspond to the unknown fields in $\ym$. To computationally solve the (either weak or strong) problem, constitutive equations must be specified, which is the subject of Section \ref{sec:constitutivetheory}.
 %
Ellipticity of operators, functional and numerical properties of the solution and of its approximation depend on the constitutive assumptions and on the choice of the correct functional spaces $\mathcal{V}^{[ 0,t_f ]}, \mathcal{V}$, whose identification falls beyond the scope of the present paper.

\subsection{Equilibrium solution}
\label{subsec:Equilibrium}

We will discriminate the equilibrium conditions, that occur at no current flowing in the electrolyte, from the steady-state conditions, in which processes simply become time-independent.
Chemical equilibrium for reaction \eqref{eq:IonizationReaction1} implies
\begin{align}
& c_{\rm Li_0}^{\rm eq} = \frac{ c^{\rm eq}_{{\rm Li}^+} }{ c^{\rm eq}_{{\rm Li}^+} + K_{\rm eq}^{\rm ion} } \, c_0
\, ,
\qquad
c^{\rm eq}_{\rm n^-}  = \frac{  K_{\rm eq}^{\rm ion} }{ c^{\rm eq}_{{\rm Li}^+} + K_{\rm eq}^{\rm ion} } \, c_0
\; .
\label{eq:ion_react_equilibrium_2}
\end{align}
In view of reaction \eqref{eq:IonizationReaction2}, part of the lithium is transformed into hopping. Hence\footnote{Note that eq. \eqref{eq:ion_react_equilibrium_3} holds because at equilibrium we assume that concentrations are uniform, thus eq. \eqref{eq:ion_react_equilibrium_3} merely expresses a mass conservation. Of course, out of equilibrium, the very same equation may describe electroneutrality, a constraint that is not imposed a priori in the present note. This point will be discussed further later on in the paper.}, {\em{at equilibrium}},
\begin{equation}
c^{\rm eq}_{{\rm Li}^+} + c^{\rm eq}_{{\rm Li}^+_{\rm hop}}  = c^{\rm eq}_{\rm n^-}
\quad
\rightarrow
c^{\rm eq}_{{\rm Li}^+_{\rm hop}}  = \frac{  K_{\rm eq}^{\rm ion} }{ c^{\rm eq}_{{\rm Li}^+} + K_{\rm eq}^{\rm ion} } \, c_0  - c^{\rm eq}_{{\rm Li}^+}
\; .
\label{eq:ion_react_equilibrium_3}
\end{equation}
Chemical equilibrium of reaction \eqref{eq:IonizationReaction2} yields
\begin{equation}
K_{\rm eq}^{\rm hop} \, c^{\rm eq}_{{\rm Li}^+} - \left(  \frac{K_{\rm eq}^{\rm ion}}{K_{\rm eq}^{\rm ion} + c^{\rm eq}_{{\rm Li}^+}} \; c_0 - c^{\rm eq}_{{\rm Li}^+}  \right) = 0
\; ,
%
\end{equation}
to be solved for $c_{{\rm Li}^+}$. It yields
\begin{equation}
{c}^{\rm eq}_{{\rm Li}^+}
=
\frac{ K_{\rm eq}^{\rm ion} }{2} \left(  \, \sqrt{ {1 + 4 \frac{c_0 }{ K_{\rm eq}^{\rm ion} } \, \frac{ 1 }{ 1 + K_{\rm eq}^{\rm hop} }} } \; - 1 \right)
\; ,
\label{eq:ion_react_equilibrium_4}
\end{equation}
to be replaced in eqs. \eqref{eq:ion_react_equilibrium_2} - \eqref{eq:ion_react_equilibrium_3}. Three independent parameters, therefore, shape the equilibrium concentrations, namely $c_0$ and the two equilibrium constant of reactions \eqref{eq:IonizationReaction1} and \eqref{eq:IonizationReaction2}. Whereas the former can be estimated with accuracy, experimental estimation of
$K_{\rm eq}^{\rm ion}$ and $K_{\rm eq}^{\rm hop}$ is subject to considerable uncertainties. The three parameters are connected to the fraction of Li that resides in equilibrium in the mobile state, termed here $\delta$ as in \cite{RaijmakersEtAlEA2020}, i.e.
\begin{align}
c_{{\rm Li_0}}^{\rm eq} = (1-\delta) \; c_0 ,
\quad
c_{{\rm n^-}}^{\rm eq} = \delta \; c_0 \; .
\label{eq:ion_react_equilibrium_5}
\end{align}
Comparing eqs. (\ref{eq:ion_react_equilibrium_2}b), \eqref{eq:ion_react_equilibrium_4}, and (\ref{eq:ion_react_equilibrium_5}b), it holds
\begin{align}
\delta \;  = \frac{  2 }{   1+ \, \sqrt{ {1 + 4 \frac{c_0 }{ K_{\rm eq}^{\rm ion} } \, \frac{ 1 }{ 1 + K_{\rm eq}^{\rm hop} }} }  }
\; .
\label{eq:keq_and_delta}
\end{align}
Equation \eqref{eq:keq_and_delta} can be easily inverted to obtain $K_{\rm eq}^{\rm hop} $ as a function of  $ \delta$ and $ K_{\rm eq}^{\rm ion} $
\begin{equation}
\label{eq:Khop_identity}
 K_{\rm eq}^{\rm hop}  = \frac{ - c_0 \delta^2  + ( 1 - \delta ) K_{\rm eq}^{\rm ion}  }{(\delta -1) K_{\rm eq}^{\rm ion} }
 \; .
\end{equation}
Since the latter can assume only positive values, eq. \eqref{eq:Khop_identity} limits the region of admissible pairs $\{  \delta ,  K_{\rm eq}^{\rm ion} \}$. 
The {\em{upper bound}} for $K_{\rm eq}^{\rm ion}$ is 
\begin{equation}
\label{eq:Keqion1}
{\overline K}_{\rm eq}^{\rm ion} = \frac{\delta^2}{1-\delta}c_0
\; ,
\end{equation}
which, by coincidence, is the equilibrium constant of reaction \eqref{eq:IonizationReaction1} defined in \cite{RaijmakersEtAlEA2020}.
In fact, a vanishing value for the equilibrium constant $K_{\rm eq}^{\rm hop}$ corresponds to ${\overline K}_{\rm eq}^{\rm ion} $ in identity \eqref{eq:keq_and_delta}. 
Moreover, negative values for $K_{\rm eq}^{\rm hop}$ come out if  ${K}_{\rm eq}^{\rm ion} >  {\overline K}_{\rm eq}^{\rm ion} $.

\section{Constitutive theory}
\label{sec:constitutivetheory}

For the sake of limiting the length of this note, we do not indulge in details on the thermodynamic balance of energy and entropy, which can be derived from \cite{SalvadoriEtAlJMPS2018} and from the appendix A in \cite{SalvadoriEtAlJPS2015}. Constitutive theory moves from the Helmholtz free energy density $\psi$ that describes the isothermal processes at hand, assumed to consist of two separate contributions:
$$
\psi( c_\alpha, \vect{E}) =
\psi_{diff}( c_\alpha ) +
\psi_{el}(\vect{E})
\; ,
$$
with $\alpha = {\rm Li^+} , {\rm Li}^+_{\rm hop}$.
The mass transport process is described by $\psi_{diff}$, adopting species concentrations $c_\alpha$ as the state variables. The contribution $\psi_{el}(\vect{E})$ models the electromagnetic interactions, in terms of the electric field $\vect{E}$. The processes are thermodynamically uncoupled.

\bigskip
The electric displacement field is related to the electric field constitutively. In linear media
\begin{equation}
\label{eq:psiel}
   \psi_{el}( \vect{E} \, ) =  - \frac{1}{2} \myvarepsilon \, \vect{E}  \cdot  \vect{E}
\end{equation}
whence, by means of identity \eqref{eq:ElectrostaticPotential},
\begin{equation}
\label{eq:elconstlaw}
  \vect{D} = - \frac{ \partial \psi_{el}( \vect{E} \, ) }{ \partial \vect{E}  } =  \myvarepsilon \; \vect{E} = -  \myvarepsilon \; \gradient{\phi}
\end{equation}
The permittivity ${\myvarepsilon = \myvarepsilon_r \, \myvarepsilon_0}$ quantifies a material's ability to transmit (or ``permit") an electric field. Its value is $8.85 \times 10^{-12} \; {\rm {C} \, V^{-1} \, m^{-1} }$ in vacuum (denoted with $ \myvarepsilon_0$) . The permittivity of a homogeneous material is usually given relative to that of vacuum, as a relative permittivity $\myvarepsilon_r$.

\bigskip
The free energy $\psi_{diff}( c_{\rm Li^+} , c_{{\rm Li}^+_{\rm hop}} )$ in a mixture, and in turn the chemical potentials
\begin{equation}
{\mu}_\alpha \, = \,  \frac{\partial \psi_{diff}( c_\alpha ) }{\partial  c_\alpha}
\; ,
\qquad
\alpha = {\rm Li^+} , {\rm Li}^+_{\rm hop}
\label{eq:chem_pot}
\end{equation}
depend on the composition of the mixture itself.
%
For no reasons but simplicity, we assume {\em{ideal}} conditions, and thus neglect the chemical interactions between ${\rm Li^+}$ and $ {\rm Li}^+_{\rm hop}$. We are aware of how strong this assumption can be, and will consider more intricate Maxwell-Stefan free energies in future works.
In order to satisfy thermodynamic consistency, see among others \cite{SalvadoriEtAlJMPS2018} and appendix A in \cite{SalvadoriEtAlJPS2015}, a linear dependence of the mass flux of species $\alpha$ on the gradient of the electrochemical potential is taken
\begin{subequations}
\begin{equation}
\label{eq:Ficksalpha}
    \vect{h}_{\alpha} = - \tensor{M}_{\alpha} \; \gradient{  {\overline \mu}_{\alpha} }
\end{equation}
by means of a positive definite mobility tensor $\tensor{M}_{\alpha}$, with the electrochemical potential ${\overline \mu}_{\alpha}$ defined as
\begin{equation}
{\overline \mu}_{\alpha}
=
\mu_\alpha + F \, z_\alpha \,  \phi
\; .
\label{eq:electrochemicalpotential}
\end{equation}
In {\em{dilute solutions far from saturation}}, the isotropic linear choice
\begin{equation}
\label{eq:isotropicmobility1nosat}
    \tensor{M}_{\alpha}( c_{\alpha}) = \ionmobility_{\alpha} \, c_{\alpha} \; \mathds{1}
\end{equation}
\label{subeq:Ficks_law}
\end{subequations}
is taken, implying that the pure phase $c_{\alpha} =0$ has a vanishing mobility. The amount $\ionmobility_{\alpha}>0 $ is usually termed the {\em{ion mobility}}. This approach is generally named after Fick's diffusion and captures an underlying brownian motion of species in a statistical sense. 

\bigskip
An {\em{ideal solution model}} \cite{DeHoffBook} provides the following free energy density for the continuum approximation of the mixing for dilute solutions far from saturation
\begin{align}
\label{eq:psidiffidealnosat}
&   \psi_{diff}^{id} (c_{\rm Li^+}, c_{{\rm Li}^+_{\rm hop}}) =    \,
	\mu_{\rm Li^+}^0  \, c_{\rm Li^+} \; +
	\mu_{{\rm Li}^+_{\rm hop}}^0  \, c_{{\rm Li}^+_{\rm hop}} \; +
       R \, \Temperature 
       \left(
        c_{ {\rm Li^+}} \, \ln[  c_{ {\rm Li^+}}  ] +
        c_{{{\rm Li}^+_{\rm hop}}} \, \ln[  c_{{{\rm Li}^+_{\rm hop}}}  ]
        \right)
        \; .
\end{align}
%
%
$R$ is the universal gas constant, $\mu_{\alpha}^0$ is a reference value of the chemical potential of diffusing species $\alpha$.
Applying \eqref{eq:chem_pot}, the chemical potential results in the form
\begin{align}
\label{eq:chempotidealnosat}
&     \mu_{\alpha}  =    \, \mu_{\alpha}^0 \; +  R \, \Temperature \,  (1+  \ln[  c_{\alpha}  ] )
\end{align}
and Fick's law \eqref{subeq:Ficks_law} takes the Nernst-Planck form
\begin{equation}
\label{eq:Nernst-Planck}
    \vect{h}_{\alpha} = -  \ionmobility_{\alpha} \, R \, \Temperature \;  \gradient{  {c}_{\alpha} } - z_{\alpha} \, F \, \ionmobility_{\alpha} \, c_{\alpha} \; \gradient{ \phi }
\end{equation}
for the flux density of species $\alpha = {\rm Li^+} , {\rm Li}^+_{\rm hop}$ in absence of convection (see for instance \cite{NewmanBook2004} ) in dilute solutions far from saturation. The {\em{diffusivity}}
%
$\diffusivity_{\alpha}$ is defined by $\diffusivity_{\alpha} = \ionmobility_{\alpha} \, R \, \Temperature$ (this equation is sometimes termed after Nernst-Einstein).

The hopping mechanism is thermodynamically quite different from the interstitial motion, thus making recourse to the classical Nernst-Planck thermodynamic description for both mechanisms might be questionable. While noting that such a form is generally accepted in the literature (see for instance \cite{Fabre2012,Danilovetal2011,DanilovNottenEA2008,RaijmakersEtAlEA2020}), we shall elaborate this issue further in future works. Within this paper, we assume that the fluxes $\vm{h}_{{\rm Li}^+}$ and $\vm{h}_{{\rm Li}^+_{\rm hop}}$ in eqs. \eqref{eq:weakmicromassbalance4}  obey the Nernst-Planck equation \eqref{eq:Nernst-Planck}.

\bigskip
Equilibrium conditions for the chemical reactions \eqref{eq:IonizationReaction1} and \eqref{eq:IonizationReaction2} can be achieved from thermodynamics, as well. They are detailed in appendix \ref{app:eqkin}.

\section{Governing equations and their weak form}
\label{sec:governingequations}

\subsection{Multiscale compatible formulation}
\label{sec:Multiscale_formulation}

 The variable fields controlling the problem result from the thermodynamic choices made, i.e. concentrations $c_{{\rm Li_0}} , \, c_{{\rm n}^-} ,  \, c_{{\rm Li}^+} , \, c_{{\rm Li}^+_{\rm hop}}$, and the electric potential $\phi$. Governing equations at all points $\vect{x} \in V$ and times $t$ come out from incorporation of the constitutive equations (\ref{eq:elconstlaw}) and  (\ref{eq:Nernst-Planck}) into the balance equations (\ref{eq:MassEle_model2}) and (\ref{eq:IncrementalGaussLaw}). They encompass eq. \eqref{eq:MassEle_model1} and the following three:
\begin{subequations}
\begin{align}
\label{eq:governingequations_a}
&  \frac{ \partial c_{{\rm Li}^+} }{ \partial t} \,- \,
  \divergence{\diffusivity_{{\rm Li}^+}  \;  \gradient{  {c}_{{\rm Li}^+} } +\, \frac{F\,\diffusivity_{{\rm Li}^+}}{RT} \, c_{{\rm Li}^+} \; \gradient{ \phi }  }=w-y\\
\label{eq:governingequations_b}
&\frac{ \partial c_{{\rm Li}^+_{\rm hop}} }{ \partial t} \,- \,
  \divergence{\diffusivity_{{\rm Li}^+_{\rm hop}}  \;  \gradient{  {c}_{{\rm Li}^+_{\rm hop}}} + \frac{F\,\diffusivity_{{\rm Li}^+_{\rm hop}}}{RT}\, c_{{\rm Li}^+_{\rm hop}} \; \gradient{ \phi } }=y\\
&
\nonumber
\divergence{  \myvarepsilon \; \gradient{ \frac{\partial {\phi }}{\partial t}   }  +
     \,   \frac{F^2}{RT} \; \left( \diffusivity_{{\rm Li}^+} \, c_{{\rm Li}^+} +  \, \diffusivity_{{\rm Li}^+_{\rm hop}} \, c_{{\rm Li}^+_{\rm hop}} \right) \; \gradient{ \phi }   } + \\
\label{eq:governingequations_c}
& \qquad
\divergence{
     F \, \left( \diffusivity_{{\rm Li}^+}  \gradient{  {c}_{{\rm Li}^+} } + \diffusivity_{{\rm Li}^+_{\rm hop}} \gradient{  {c}_{{\rm Li}^+_{\rm hop}} }  \right) } = 0
\end{align}
\label{eq:governingequations}
\end{subequations}
together with mass action laws \eqref{eq:mass_action_ion}, \eqref{eq:mass_action_hop}.

It is typical in batteries to fully impose Neumann conditions for concentration, in terms of mass fluxes, during galvanostatic processes. Accordingly, conditions (\ref{eq:MassFluxBC}) and (\ref{eq:ElPotBC}) are applied along Neumann boundaries $\partial^N V$. To ensure uniqueness, Dirichlet boundary conditions have to be imposed along part $\partial^D V$, being $\partial V=\partial^D V \cup \partial^N V$.

Initial conditions usually enforce equilibrium. They have been stated in equations \eqref{eq:ion_react_equilibrium_2}-\eqref{eq:ion_react_equilibrium_4}.
Initial conditions for electric potential should match the boundary value problem at $t=0$, and yield a uniform value at all $\vect{x} \in V$. At initial time, in fact, Gauss law (\ref{eq:GaussLaw}) provides the necessary and sufficient equations to be solved for $\phi$:
\begin{equation}
\divergence{ - \myvarepsilon \; \gradient{ \phi  }   } =  F \, \left(  c_{{\rm Li}^+} + c_{{\rm Li}^+_{\rm hop}} -  c_{\rm n^-} \right)  \qquad {\vect{x}} \in V, \; t=0
\label{eq:initcondeq}
\end{equation}
together with homogeneous boundary conditions for potential and current - in view of thermodynamic equilibrium at initial time.

\bigskip
The evolution problem can be formulated in a weak form as well. Following a Galerkin approach, weak forms are built {\em{at a given time t}} using ``variations'' of the same set of variables that rule the problem, namely concentrations ${\hat c}_\alpha$ and electric potential ${\hat \phi}$ which are solely depending upon the space variable $\vect{x}$. The weak form of the {\em{governing}} equations can thus be written in time interval $\left[ 0, t_f \right]$ as
\begin{subequations}
\begin{align}
\nonumber
& \mbox{ Find } \ym \in \mathcal{V}^{[ 0,t_f ]}  \mbox{ such that }
\\
&
\hspace{1cm}
 \frac{\rm d}{ {\rm d}t} b \left( \ymh({\vect{x}}),  {y}({\vect{x}},t) \right) + a ( \ymh({\vect{x}}),  {y}({\vect{x}},t) ) + c ( \ymh({\vect{x}}),  {y}({\vect{x}},t) )= f(\ymh({\vect{x}}))
\hspace{1cm}
 \forall \ymh \in \mathcal{V}
\end{align}
where
\begin{align}
b \left( \ymh, {z} \right) &=
   \int_{V} \;   {\hat { c}}_{{\rm Li_0}} \, c_{{\rm Li_0}} +{\hat { c}}_{{\rm n}^{\!-}} \, c_{{\rm n}^-}+{\hat { c}}_{{\rm Li}^+} \, c_{{\rm Li}^+}+{\hat { c}}_{{\rm Li}^+_{\rm hop}} \, c_{{\rm Li}^+_{\rm hop}} \,	+
   \;  \myvarepsilon \; \gradient{ {\hat \phi}} \cdot \gradient{ \frac{\partial {\phi }}{\partial t}   } \, {\rm d}V
 \\ \nonumber
a \left( \ymh, {z}(t) \right)  &=
     \int_{V}\diffusivity_{{\rm Li}^+}\gradient{ {\hat c}_{{\rm Li}^+} } \cdot    \gradient{  {c}_{{\rm Li}^+} } + \frac{F\diffusivity_{{\rm Li}^+}}{RT} c_{{\rm Li}^+}  \gradient{ {\hat c}_{{\rm Li}^+} } \cdot  \gradient{ \phi }{\rm d}V+
    \\ \nonumber
    & + \int_{V}\diffusivity_{{\rm Li}^+_{\rm hop}}\gradient{ {\hat c}_{{\rm Li}^+_{\rm hop}} } \cdot    \gradient{  {c}_{{\rm Li}^+_{\rm hop}} } + \frac{F\diffusivity_{{\rm Li}^+_{\rm hop}}}{RT} \, c_{{\rm Li}^+_{\rm hop}}  \gradient{ {\hat c}_{{\rm Li}^+_{\rm hop}} } \cdot  \gradient{ \phi } {\rm d}V+
    \\ \nonumber
    &+ \int_{V} F  \; \gradient{ {\hat \phi}} \cdot \left( \diffusivity_{{\rm Li}^+}  \;  \gradient{  {c}_{{\rm Li}^+} }+ \diffusivity_{{\rm Li}^+_{\rm hop}}  \;  \gradient{  {c}_{{\rm Li}^+_{\rm hop}} } \, \right) \;  {\rm d}V
  \\
  &+\int_{V} \frac{F^2}{RT} \, \left( \diffusivity_{{\rm Li}^+} \, c_{{\rm Li}^+} +  \, \diffusivity_{{\rm Li}^+_{\rm hop}} \, c_{{\rm Li}^+_{\rm hop}} \right) \; \gradient{ {\hat \phi}} \cdot \gradient{ \phi }\;  {\rm d}V
\\
c \left( \ymh, {z}(t) \right)  &=
    \int_{V} \;{\hat { c}}_{{\rm Li_0}}\cdot w - \;{\hat { c}}_{{\rm n}^-}\cdot w \, - \;{\hat { c}}_{{\rm Li}^+}\cdot (w-y) \, - \;{\hat { c}}_{{\rm Li}^+_{\rm hop}}\cdot y\,{\rm d}V
\\
f \left( \ymh \right) &=
 - \int_{\partial^N V } \; {\hat { c}}_{{\rm Li}^+} {h}^{BV}_{{\rm Li}^+}+{\hat { c}}_{{\rm Li}^+_{\rm hop}} {h}^{BV}_{{\rm Li}^+_{\rm hop}}-F \hat{ \phi}\left ({h}^{BV}_{{\rm Li}^+}+{h}^{BV}_{{\rm Li}^+_{\rm hop}}\right) \; {\rm d} \Gamma	
\end{align}
 \label{eq:gov_eqs_weakform}
\end{subequations}
with ${y}({\vect{x}},t)= \{\, c_{{\rm Li_0}} , \, c_{{\rm n}^-} ,  \, c_{{\rm Li}^+} , \, c_{{\rm Li}^+_{\rm hop}},\phi \}$ and  $\ymh({\vect{x}}) = \{ {\hat c}_{{\rm Li_0}} , \, {\hat c}_{{\rm n}^-} ,  \, {\hat c}_{{\rm Li}^+} , \, {\hat c}_{{\rm Li}^+_{\rm hop}}, {\hat \phi} \}$.

\subsection{Approximated electroneutral formulation}
\label{sec:Electroneutral formulation}

The hypothesis of electroneutrality, namely equation \eqref{eq:electroneutrality}, leads to a simpler formulation for the governing equations. They encompass eq. \eqref{eq:electroneutrality} itself, eq. \eqref{eq:MassEle_model1}, eq. \eqref{eq:governingequations_a} and a linear combination of the former with \eqref{eq:governingequations_b}, which eventually leads to:
\begin{align}
 \divergence{
 \diffusivity_{{\rm Li}^+}  \;  \gradient{  {c}_{{\rm Li}^+} }
 +  \diffusivity_{{\rm Li}^+_{\rm hop}}  \;  \gradient{  {c}_{{\rm Li}^+_{\rm hop}}}
 + \, \frac{F}{RT}
 \left(
  \diffusivity_{{\rm Li}^+} \, c_{{\rm Li}^+}
 +
 \diffusivity_{{\rm Li}^+_{\rm hop}} \, c_{{\rm Li}^+_{\rm hop}}
 \right)
 \; \gradient{ \phi }
 } =0
\label{eq:governingequations_electro}
\end{align}

\section{Steady state solution}
\label{sec:Steady state solution}

At the end of the transient behavior, the solid electrolyte response will reach a steady state at which the fields $c_{{\rm Li_0}} , \, c_{{\rm n}^-} ,  \, c_{{\rm Li}^+} , \, c_{{\rm Li}^+_{\rm hop}}$, and $\phi$ do not change in time.
Note that at steady state current flows in the electrolyte, hence the latter is still out of equilibrium.

A closed form solution at steady state can be found for one-dimensional systems under the assumption of electroneutrality. We may start from eq. \eqref{eq:MassEle_model1}. Since the left hand side must vanish, then $w=0$, i.e. the ionization reaction \eqref{eq:IonizationReaction1} must be at equilibrium. Hence, conditions \eqref{eq:ion_react_equilibrium_2} hold, here copied and pasted for readability
\begin{align}
& c_{\rm Li_0}^{\rm ss} = \frac{ c^{\rm ss}_{{\rm Li}^+} }{ c^{\rm ss}_{{\rm Li}^+} + K_{\rm eq}^{\rm ion} } \, c_0
\, ,
\qquad
c^{\rm ss}_{\rm n^-}  = \frac{  K_{\rm eq}^{\rm ion} }{ c^{\rm ss}_{{\rm Li}^+} + K_{\rm eq}^{\rm ion} } \, c_0
\; .
\label{eq:ion_react_equilibrium_2_copy}
\end{align}
with the apex ``$\rm ss$" that stands for steady state. Enforcing electroneutrality, eq. \eqref{eq:electroneutrality} implies
\begin{subequations}
\begin{align}
&  c^{\rm ss}_{{\rm Li}^+_{\rm hop}}=\frac{K_{\rm eq}^{\rm ion}}{K_{\rm eq}^{\rm ion} + c^{\rm ss}_{{\rm Li}^+}} \, c_0  -  c^{\rm ss}_{{\rm Li}^+}.
\end{align}
\label{eq:steady-state}
\end{subequations}
We are thus left with two unknown fields $c^{\rm ss}_{{\rm Li}^+}$, $\phi^{\rm ss}$. They can be found solving eqs. \eqref{eq:governingequations_b} and \eqref{eq:governingequations_electro}, i.e.
\begin{subequations}
\begin{align}
\label{eq:governingequationsSteady1_a}
&-\divergence{\gradient{c^{\rm ss}_{{\rm Li}^+_{\rm hop}}} +\, \frac{F}{RT} \, c^{\rm ss}_{{\rm Li}^+_{\rm hop}} \; \gradient{ \phi^{\rm ss} }  }=-\frac{y}{\diffusivity_{{\rm Li}^+}},\\
\label{eq:governingequationsSteady1_b}
&-\divergence{\gradient{\frac{K_{\rm eq}^{\rm ion}}{K_{\rm eq}^{\rm ion}+c^{\rm ss}_{{\rm Li}^+}}c_0} +\, \frac{F}{RT} \, \left(\frac{K_{\rm eq}^{\rm ion}}{K_{\rm eq}^{\rm ion}+c^{\rm ss}_{{\rm Li}^+}}c_0\right) \; \gradient{ \phi^{\rm ss} }  }=\left(\frac{1}{\diffusivity_{{\rm Li}^+_{\rm hop}}}-\frac{1}{\diffusivity_{{\rm Li}^+}}\right) y,
\end{align}
\label{eq:governingequationsSteady1}
\end{subequations}
with $y$ as in \eqref{eq:mass_action_hop}.
A uniform concentration $c^{\rm ss}_{{\rm Li}^+}(x)=\overline{c}_{{\rm Li}^+}$ is sought for. In this circumstance,
summing up the two equations \eqref{eq:governingequationsSteady1_a}, \eqref{eq:governingequationsSteady1_b}
and rearranging, the laplacian of the electric potential turns out to be defined through an unknown constant $p$ as follows
\begin{equation}
\divergence{\gradient{ \phi }} 
=
\frac{1}{\diffusivity_{{\rm Li}^+}}
\,
\frac{k_f^{\rm hop} \; \overline{c}_{{\rm Li}^+}-k_b^{\rm hop}\left(\frac{K_{\rm eq}^{\rm ion}}{K_{\rm eq}^{\rm ion}+\overline{c}_{{\rm Li}^+}}c_0-\overline{c}_{{\rm Li}^+}\right)}
       {\frac{F}{RT}\left(\overline{c}_{{\rm Li}^+}-\frac{K_{\rm eq}^{\rm ion}}{K_{\rm eq}^{\rm ion}+\overline{c}_{{\rm Li}^+}}c_0\right)} 
=
p \; e^{   \frac{ \zeta F \phi}{RT} }
\; ,
\label{eq:governingequationsSteady4}
\end{equation}
taking advantage of eq. \eqref{eq:kasfunctionofphi}.
Note that the numerator in eq.\eqref{eq:governingequationsSteady4} is equal to $y$ from eq.\eqref{eq:mass_action_hop}.
Restricting to a one-dimensional problem, the solution of \eqref{eq:governingequationsSteady4} is 
$$
 \phi =  a+b x +  \frac{ R^2 T^2  }{  \zeta^2 F^2 } \, p \; e^{   \frac{ \zeta F \phi}{RT} }  
$$
Imposing $\phi(0) = 0$ and the conservation of current $\vect{h}_{{\rm Li}^+}(L)+\vect{h}_{{\rm Li}^+_{\rm hop}}(L)=\vect{h}_{{\rm Li}^+}(0)+\vect{h}_{{\rm Li}^+_{\rm hop}}(0)$,
the constant $p$ must vanish, and hence $y$ must vanish, too, hence chemical equilibrium is granted.

We conclude that, at steady state, the equilibrium concentration \eqref{eq:ion_react_equilibrium_4} holds for ${c}^{\rm ss}_{{\rm Li}^+}$ and the electric potential is linear 
\begin{equation}
\label{eq:steadystatephi}
\phi= b \, x 
\; ,
\end{equation}
with constant $b$ identified by the current $\vect{h}_{{\rm Li}^+}(L)+\vect{h}_{{\rm Li}^+_{\rm hop}}(L)$ that flows across the electrolyte. 
Since the equilibrium concentrations are uniform, Nernst-Planck constitutive equation \eqref{eq:Nernst-Planck} has vanishing diffusive contribution and reduces to a special form of Ohm's law, with conductivity that depends upon concentration. From Faraday's law \eqref{eq:FaradaysLaw}, it descends
\begin{equation}
\label{eq:Nernst-Planck_ss}
    \vect{i}=  - \, \frac{F^2}{RT}
 \left(
  \diffusivity_{{\rm Li}^+} \, c^{\rm ss}_{{\rm Li}^+}
 +
 \diffusivity_{{\rm Li}^+_{\rm hop}} \, c^{\rm ss}_{{\rm Li}^+_{\rm hop}}
 \right)
 \; \gradient{ \phi }
 \; .
\end{equation}

\section{Numerical simulations}
\label{sec:numerical simulation}

To validate the model described so far, the {\em{one-dimensional}} solid electrolyte case study of \cite{RaijmakersEtAlEA2020} will be analyzed. The response of the electrolyte, part of a commercial all-solid-state thin-film battery with storage capacity of $0.7 \rm mAh $, is simulated under galvanostatic conditions of charge, at a constant temperature of $25^o \rm C$ and zero state of stress.
The electrolyte has been experimentally tested in \cite{RaijmakersEtAlEA2020} at different $C\!-\!rates$.
%
%
 All the parameters shared with \cite{RaijmakersEtAlEA2020} have been taken from that paper.
 %

The electrolyte is a layer of LiPON with thickness 3.62 $\rm \mu m$. 
We take for the ${\rm Li}^+_{\rm hop}$ diffusivity the vacancy diffusivity provided in \cite{RaijmakersEtAlEA2020}, i.e.  $\diffusivity_{{\rm Li^+}_{\rm hop}}=5.69 \times 10^{-16} \rm m^2 s^{-1}$, whereas  the interstitial lithium diffusivity holds $\diffusivity_{{\rm Li}^+}=1.73 \times 10^{-16} \rm m^2 s^{-1}$. The relative permittivity of LiPON is assumed in the range of $1-100$.
The fraction of Li that resides in equilibrium in the mobile state, $\delta$ is taken as in \cite{RaijmakersEtAlEA2020}, i.e. $\delta = 0.64$.

A few test-cases have been run changing the reaction constants parameters.

%
%

%
\redcom{

%
\begin{table}[!htb]
\small
\centering
\begin{tabular}{ |p{1.70cm}||p{1.70cm}|p{2.5cm}|p{8.90cm}|}
 \hline
 \multicolumn{4}{|c|}{Input parameters} \\
 \hline
 Parameters      & Value                         & Unit of measure       & Description \\
 \hline
 $T$                          &  $298.5$                & $°K$          &Temperature\\
 $L$                          &  $3.62\cdot10^{-6}$     & $m$           &Thickness of the electrolyte\\
 $A$                          &  $3.36\cdot10^{-4}$     & $m^2$         &Geometrical surface area\\
 $\diffusivity_{{\rm Li}^+}$                   &  $5.69\cdot10^{-16}$    &$m^2/s$        &Diffusion coefficient for $\rm Li^+$ ions in the electrolyte\\
 $\diffusivity_{{\rm Li}^+_{\rm hop}}$             &  $1.73\cdot10^{-16}$    &$m^2/s$        &Diffusion coefficient for $\rm Li^+_{hop}$ in the electrolyte\\
 $\delta$                     &  $0.64$    &-              &Fraction of mobile $\rm Li$ in the electrolyte in equilibrium\\
 $c_0$                        &  $61141$                &$mol/m^3$      &Maximal lithium concentration in the electrolyte\\
  \hline
\end{tabular}
\caption{\em{Model parameters used during simulations. }}
\label{Tab:Input}
\end{table}
}

In order to make initial and boundary conditions compatible with thermodynamic equilibrium at $t=0$, the current $I(t)$ is tuned in time as
\begin{equation}
 I(t) = ( 1-e^{-t} ) \, I_{1C}
 \label{eq:iTune}
\end{equation}
with $t$ in seconds and $I_{1C}$ the current at 1C-rate, i.e. $I_{1C}=0.70 \rm mA$.
\begin{figure}[!htb]
\begin{center}
\includegraphics[ width=12cm ]{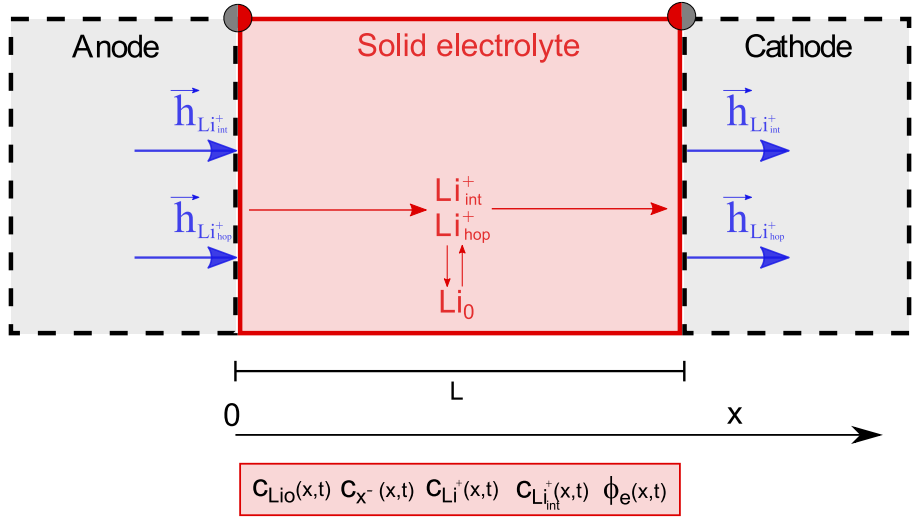}
\end{center}
\caption{\it A one-dimensional model of a Li-ion battery, with separator of size $3.62\cdot10^{-6}$ $\rm m$ highlighted. The flux of ${\rm Li^+}$ ions during charge is pointed out. }
\label{fig:BatteryScheme}
\end{figure}
Boundary and initial conditions have been taken according to \cite{RaijmakersEtAlEA2020}, adopting the coordinate system depicted in Fig. \ref{fig:BatteryScheme}. Initially (at $t = 0$) the concentration of ions across the electrolyte is uniform and at equilibrium as in equations \eqref{eq:ion_react_equilibrium_2}-\eqref{eq:ion_react_equilibrium_4}.
%
%
The current passing the interfaces electrode/electrolyte is the sum of the currents due to interstitial and hopping lithium. Boundary conditions (\ref{eq:MassFluxBC}) thus shall satisfy the constraints
\begin{subequations}
\begin{align}
\left( {\vect{h}}_{{\rm Li}^+}(0,t)+{\vect{h}}_{{\rm Li}^+_{\rm hop}}(0,t) \right) \cdot {\vect{n}} =- \frac{I(t)}{F A},\\
\left( {\vect{h}}_{{\rm Li}^+}(L,t)+{\vect{h}}_{{\rm Li}^+_{\rm hop}}(L,t) \right) \cdot {\vect{n}}  = \frac{I(t)}{F A},
\end{align}
\label{eq:pb1bc}
\end{subequations}
where $A =3.36 \times 10^{-4} \rm m^2 $ is the net area of the electrodes/electrolyte interfaces. For being in thermodynamic equilibrium with neither current nor mass flowing, the electric potential at initial time satisfies equations (\ref{eq:initcondeq}) and has to be homogeneous
\begin{equation}
 \phi( x, 0) = 0
 \qquad {\vect{x}} \in V
 \, .
 \label{eq:initfipb1}
\end{equation}
The boundary condition for $\phi$ at $x=0$ is homogeneous, too.

\subsection{Discretization and time advancing by finite differences}
\label{sec:discretizationtimeadvancing1D}

The weak form (\ref{eq:gov_eqs_weakform}) can be transformed in a first order Ordinary Differential Equation (ODE) in time if discretization is performed via separated variables, with spatial test $\varphi_i(x)$ and shape functions $\varphi_j(x)$ and nodal unknowns (collectively gathered in column $y$ with component ${y}_j(t)$) that depend solely on time.
The usual Einstein summation convention is taken henceforth for repeated indexes.
The non linear ODE reads:
\begin{subequations}
\begin{equation}
\mbox{ Find } {y}(t)
\mbox{ s.t. }
\qquad
 b_{i} \, \cdot  \, \frac{\partial {y}}{\partial t}(t) +  a_i [ \; {y}(t) \; ] +  c_i [ \; {y}(t) \; ] = f_i(t)
\qquad
\mbox{for } i= 1, 2, ..., N
 \label{eq:discreteweakformpb1}
\end{equation}
where
\begin{align}
b_{i} \, \cdot  \, \frac{\partial {y}}{\partial t}(t)  &=
   \int_0^l   \,
     \varphi_i^{{\rm Li_0}} \;  \varphi_j^{{\rm Li_0}}
    \,   {\rm d}x
    \;
     \frac{\partial {c}_j^{{\rm Li_0}} }{\partial t}\,
 +
  \int_0^l   \,
    \varphi_i^{{\rm n}^-} \;  \varphi_j^{{\rm n}^-}
    \,   {\rm d}x
    \;
         \frac{\partial c_j^{{\rm n}^-}}{\partial t}\,
 +
 \int_0^l   \,
     \varphi_i^{{\rm Li}^+} \;  \varphi_j^{{\rm Li}^+}
    \,   {\rm d}x
    \;
     \frac{\partial {c}_j^{{\rm Li}^+} }{\partial t}\,
 +\\ \nonumber
 &+ \int_0^l   \,
     \varphi_i^{{\rm Li}^+_{\rm hop}} \;  \varphi_j^{{\rm Li}^+_{\rm hop}}
    \,   {\rm d}x
    \;
     \frac{\partial {c}_j^{{\rm Li}^+_{\rm hop}} }{\partial t}\,
 +
 \;  \int_0^l \,\myvarepsilon \;
  \gradientone{ \varphi_i^{\phi} } \, \gradientone{ \varphi_j^{\phi}  }
  \,  {\rm d}x \,
       \frac{\partial \phi_j}{\partial t}\,
%
 \\
a_i [ \; {y}(t) \; ]  &=
   \, \;  \int_0^l \,\diffusivity_{{\rm Li}^+}
     \gradientone{ \varphi_i^{{\rm Li}^+} } \,    \gradientone{  \varphi_j^{{\rm Li}^+}  }
  \, {\rm d}x
  \; {c}_j^{{\rm Li}^+} \,
  +
 \, \;  \int_0^l   \,\diffusivity_{{\rm Li}^+_{\rm hop}}
     \gradientone{ \varphi_i^{{\rm Li}^+_{\rm hop}} } \,    \gradientone{  \varphi_j^{{\rm Li}^+_{\rm hop}}  }
  \, {\rm d}x
  \; {c}_j^{{\rm Li}^+_{\rm hop}}
  \\ \nonumber
  &-F\, \;  \int_0^l   \,
     \gradientone{ \varphi_i^{\phi} } \, \diffusivity_{{\rm Li}^+} \gradientone{  \varphi_j^{{\rm Li}^+}  }
  \, {\rm d}x  \; {c}_j^{{\rm Li}^+}\,-F\, \;  \int_0^l   \,
     \gradientone{ \varphi_i^{\phi} } \, \diffusivity_{{\rm Li}^+_{\rm hop}} \gradientone{  \varphi_k^{{\rm Li}^+_{\rm hop}}  }
  \, {\rm d}x   \; {c}_k^{{\rm Li}^+_{\rm hop}}\,
    +\\ \nonumber
 &+
 \, \frac{F}{RT}\int_0^l\,\diffusivity_{{\rm Li}^+}
  \,\gradientone{ \varphi_i^{{\rm Li}^+} } \, \varphi_j^{{\rm Li}^+} \;    \gradientone{ \varphi_k^{\phi} }
  \, {\rm d}x
  \; {c}_j^{{\rm Li}^+} \, \phi_k \;
 +
 \\ \nonumber
 & +
  \,\frac{F}{RT}\int_0^l   \,\diffusivity_{{\rm Li}^+_{\rm hop}}
  \,\gradientone{ \varphi_i^{{\rm Li}^+_{\rm hop}} } \, \varphi_j^{{\rm Li}^+_{\rm hop}} \;    \gradientone{ \varphi_k^{\phi} }
  \, {\rm d}x
  \; {c}_j^{{\rm Li}^+_{\rm hop}} \, \phi_k \;
 +
 \\ \nonumber
 & -\frac{F^2}{RT}
 \, \;  \int_0^l  \,\diffusivity_{{\rm Li}^+}
     \gradientone{ \varphi_i^{\phi}} \, \gradientone{ \varphi_k^{\phi}  }  \, \varphi_j^{{\rm Li}^+}
 \;  {\rm d}x
 \; {c}_j^{{\rm Li}^+} \, \phi_k \;
-\frac{F^2}{RT}
 \, \;  \int_0^l  \,\diffusivity_{{\rm Li}^+_{\rm hop}}
     \gradientone{ \varphi_i^{\phi}} \, \gradientone{ \varphi_k^{\phi}  }  \, \varphi_j^{{\rm Li}^+_{\rm hop}}
 \;  {\rm d}x
 \; {c}_j^{{\rm Li}^+_{\rm hop}} \, \phi_k \;
\\
c_i [ \; {y}(t) \; ]  &=
    \int_0^l \;\varphi_i^{{\rm Li_0}}\cdot (k_f^{\rm ion}\,\varphi_j^{\rm Li_0}\,c_j^{\rm Li_0}-k_b^{\rm ion}\,\varphi_j^{{\rm Li}^+}\,c_j^{{\rm Li}^+}\,\varphi_k^{\rm n^-}\,c_k^{\rm n^-})\,{\rm d}x+\\ \nonumber
    &-\int_0^l \;\varphi_i^{{\rm n}^-}\cdot (k_f^{\rm ion}\,\varphi_j^{\rm Li_0}\,c_j^{\rm Li_0}-k_b^{\rm ion}\,\varphi_j^{{\rm Li}^+}\,c_j^{{\rm Li}^+}\,\varphi_k^{\rm n^-}\,c_k^{\rm n^-}){\rm d}x\\ \nonumber
    &-\int_0^l \;\varphi_i^{{\rm Li}^+}\cdot\left [(k_f^{\rm ion}\,\varphi_j^{\rm Li_0}\,c_j^{\rm Li_0}-k_b^{\rm ion}\,\varphi_j^{{\rm Li}^+}\,c_j^{{\rm Li}^+}\,\varphi_k^{\rm n^-}\,c_k^{\rm n^-})  \right . + \\ \nonumber
    & \qquad \qquad \qquad \qquad \qquad \left.-(k_f^{\rm hop}\,\varphi_j^{{\rm Li}^+}\,c_j^{{\rm Li}^+}-k_b^{\rm hop}\,\varphi_j^{{\rm Li}^+_{\rm hop}}\,c_j^{{\rm Li}^+_{\rm hop}})\right ]\,{\rm d}x+\\ \nonumber
    &-\int_0^l \;\varphi_i^{{\rm Li}^+_{\rm hop}}\cdot (k_f^{\rm hop}\,\varphi_j^{{\rm Li}^+}\,c_j^{{\rm Li}^+}-k_b^{\rm hop}\,\varphi_j^{{\rm Li}^+_{\rm hop}}\,c_j^{{\rm Li}^+_{\rm hop}})\,{\rm d}x
\\
f_i(t)\, & = \,
 \int_{\partial^N V }  \; \varphi_i^{{\rm Li}^+} {h}^{BV}_{{\rm Li}^+} \,+ \varphi_i^{{\rm Li}^+_{\rm hop}} {h}^{BV}_{{\rm Li}^+_{\rm hop}}-F \varphi_i^{{\phi}}\left ({h}^{BV}_{{\rm Li}^+}+{h}^{BV}_{{\rm Li}^+_{\rm hop}}\right) \; {\rm d} \Gamma
\end{align}
\end{subequations}
with  $y_j(t) = \{  { c}_j^{{\rm Li_0}},  { c}_j^{{\rm n}^-},{ c}_j^{{\rm Li}^+}, { c}_j^{{\rm Li}^+_{\rm hop}},  \phi_j\}$.
A family of time-advancing methods based on the so-called $\theta$-scheme can be set up for the discrete problem (\ref{eq:discreteweakformpb1}). Assume that solution $y( t )$ is given at time $t$, and that the algorithm is triggered at the initial time $t=0$ by means of initial conditions.  The scheme seeks for ${y}( t+ \Delta t )$ such that
\begin{align}
 \nonumber
 & b_{i}  \, \cdot  \, \frac{ y( t+ \Delta t ) - y( t ) }{\Delta t} +  a_i [ \;  \theta \, y( t+ \Delta t )  + (1-\theta) y( t )  \; ] +  c_i [ \;  \theta \, y( t+ \Delta t )  + (1-\theta) y( t )  \; ]
 \\ & \qquad \qquad \qquad \qquad
 = \;  \theta \, f_i( t+ \Delta t )  + (1-\theta) f_i( t )
 \label{eq:Thetadiscreteweakformpb1}
\end{align}
for $ i= 1, 2, ..., N$, where $0 \le \theta \le 1$, $\Delta t = t_f/N_t$ is the time step, $N_t$ is a positive integer.  $\theta$-scheme includes the forward Euler scheme ($\theta = 0$, linear in $ y( t+ \Delta t ) $), backward Euler  ($\theta = 1$),  and Crank-Nicholson ($\theta = 1/2$).
In the numerical simulations that follows, backward Euler  ($\theta = 1$) has been selected, thus searching for ${y}( t+ \Delta t )$ such that
\begin{equation}
 b_{i}  \, \cdot  \, \frac{ y( t+ \Delta t ) }{\Delta t}  +  a_i [  \; y( t+ \Delta t )  \; ] +  c_i [  \; y( t+ \Delta t )  \; ]= \;  f_i( t+ \Delta t )  + b_{i}  \, \cdot  \, \frac{y( t ) }{\Delta t}
 \label{eq:BEdiscreteweakformpb1}
\end{equation}
%
Denoting with ${\sf q}=1,2,...$ the iteration counter, the Newton Raphson iterative solution scheme solves non-linear problem (\ref{eq:BEdiscreteweakformpb1}).  It proceeds until a condition on the ${\rm L}_2$ norm of the increment
$$\delta y={^{\sf q+1}}y( t+ \Delta t )-{^{\sf q}}y( t+ \Delta t )$$
is satisfied.
%
The numerical technique has been implemented in a Matlab package script, provided solutions with $20$ coincident digits and converged up to a tolerance of $10^{-10}$ in the ${\rm L}_2$ norm of the (relative) increment.

\bigskip
Several simulations have been carried out with different time steps and number of elements in order to check convergence, but those details will not be presented here.
The outcomes reported henceforth refer to a spatial discretization of the thickness of the electrolyte by means of $20$ finite elements, biased to a finer mesh in proximity of the boundaries, where greater gradient of the variables are expected. Time discretization is achieved with a constant time step $\Delta t =1$ s. 

Since the permittivity is extremely small, instabilities and convergence issue may arise. To this aim, the solution scheme has been partitioned into two separated algorithms. At first, the electroneutral approximation has been taken and the problem depicted in section \ref{sec:Electroneutral formulation} has been solved. Such a solution is used as initial guess for the multiscale compatible formulation described in section \ref{sec:Multiscale_formulation}. Although our analysis completely lacks of profound numerical analysis investigation, no problem of stability and convergence further arose in the implementation of quasi electrostatic Maxwell's equation. 

\bigskip \noindent
\subsection{Steady state response} 
\label{subsec:Steady_state_response}
\begin{figure}[htbp]
\centering
\begin{subfigure} {0.495\textwidth}
  \includegraphics[width=90mm]{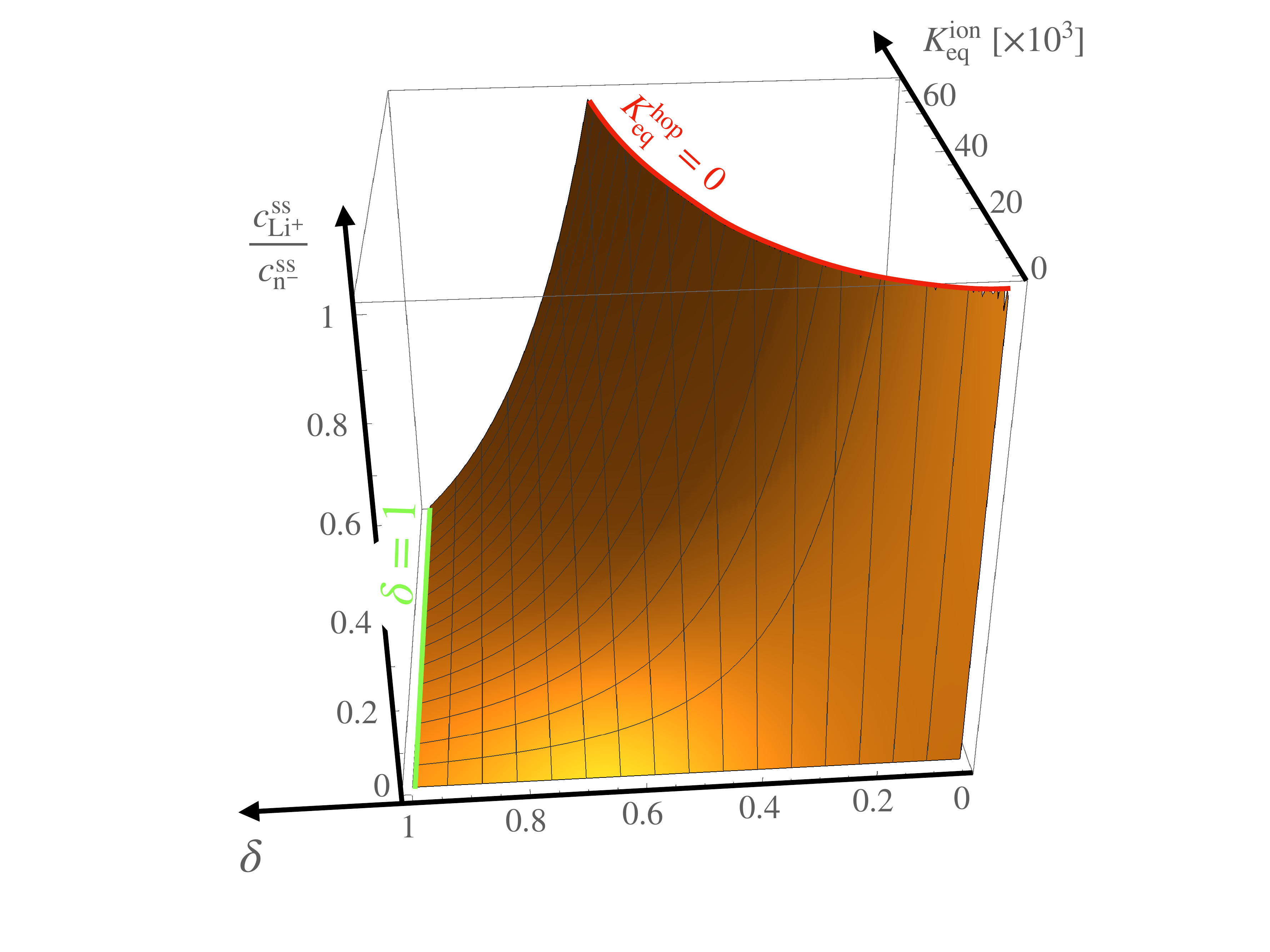}
\caption{$c_{{\rm Li}^+}$}
\label{subfig:cliss}
\end{subfigure}
\begin{subfigure} {0.495\textwidth}
  \includegraphics[width=90mm]{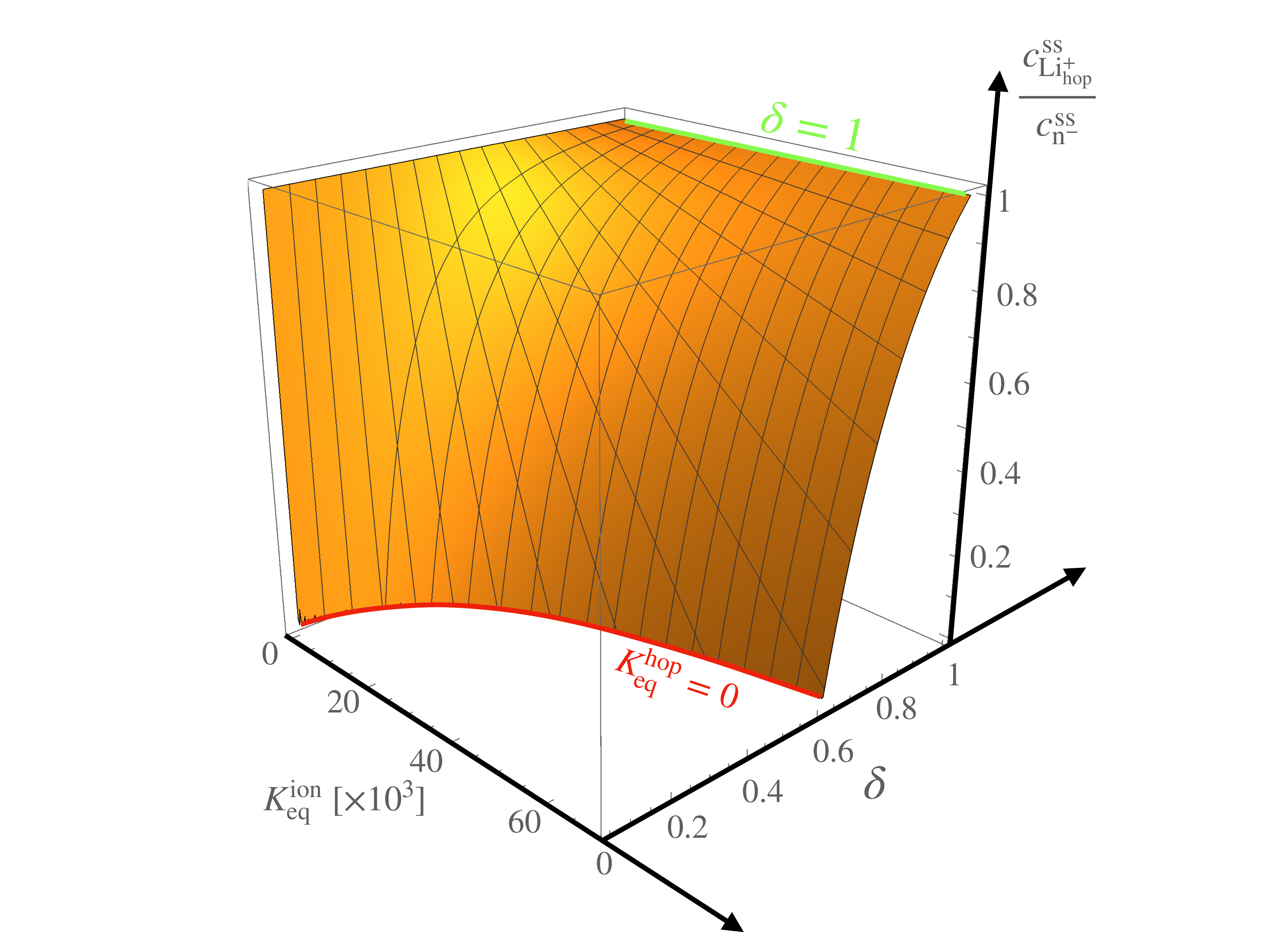}
\caption{$c_{{\rm Li}^+_{\rm hop}}$}
\label{subfig:clihopss}
\end{subfigure}
\caption{ \it Steady state representation of interstitial $c_{{\rm Li}^+}^{\rm ss}$ and hopping $c_{{\rm Li}^+_{\rm hop}}^{\rm ss}$ lithium normalized by $c_{n^-}^{\rm ss}$ as a function of $\delta$ and $K_{\rm eq}^{\rm ion}$.
The red curve  corresponds to the upper bound ${\overline K}_{\rm eq}^{\rm ion} $ defined in \eqref{eq:Keqion1},
imposing constraint $ K_{\rm eq}^{\rm hop}  = 0$ in identity \eqref{eq:Khop_identity}.}
\label{fig:cLiss_draw}
\end{figure}

The steady state response of the system was discussed in Sec. \ref{sec:Steady state solution} and is here elaborated using material parameters from Table \ref{Tab:Input}. Since the fraction of Li that resides in equilibrium in the mobile state is given, $\delta=0.64$, the values at steady state for $c_{{\rm Li_0}}^{\rm ss} $ and $c_{{\rm n^-}}^{\rm ss}$ emanate from eq. \eqref{eq:ion_react_equilibrium_5}, because steady state implies chemical equilibrium. Easy algebra allows relating the steady state values for interstitial and hopping lithium to $\delta$ and $K_{\rm eq}^{\rm ion}$:
\begin{align}
c_{{\rm Li}^+}^{\rm ss} = K_{\rm eq}^{\rm ion} \left( \frac{1}{\delta} -1 \right) \; ,
\quad
c_{{\rm Li}^+_{\rm hop}}^{\rm ss} = \delta \; c_0 + K_{\rm eq}^{\rm ion} \left(1- \frac{1}{\delta} \right) 
\; .
\label{eq:steadystatedelta}
\end{align}
Figure \ref{fig:cLiss_draw} plots the evolution of $c_{{\rm Li}^+}^{\rm ss}$ and $c_{{\rm Li}^+_{\rm hop}}^{\rm ss}$ normalized by the concentration of vacancies $c_{{\rm n}^-}^{\rm ss}$ at different values for $K_{\rm eq}^{\rm ion}$ and $\delta$. 
The red curve in the plot corresponds to the upper bound ${\overline K}_{\rm eq}^{\rm ion} $ defined in \eqref{eq:Keqion1},
which imposes constraint $ K_{\rm eq}^{\rm hop}  = 0$  in identity \eqref{eq:Khop_identity}. Such a curve, in the $\{ \delta, K_{\rm eq}^{\rm ion} \}$ plane, emerges at $c_{{\rm Li^+}}^{\rm ss}=c_{{\rm n^-}}^{\rm ss}$, as expected since no hopping takes place. Two further curves are of interest in Fig. \ref{fig:cLiss_draw}. At $K_{\rm eq}^{\rm ion} = 0$ one sees that $c_{{\rm Li^+}}^{\rm ss}$ vanishes. This outcome is largely predicted, since no ions are generated in the chemical ionization reaction \eqref{eq:IonizationReaction1}. On the other hand, as $\delta$ approaches the limit unit value, $c_{{\rm Li^+}}^{\rm ss}$ tends to zero again. In such a case, the chemical ionization reaction \eqref{eq:IonizationReaction1} is complete and all host-sites became negative vacancies. No matter the value for $K_{\rm eq}^{\rm ion}$,  $K_{\rm eq}^{\rm hop }$ becomes larger and larger (see eq. \eqref{eq:Khop_identity}) hence also reaction \eqref{eq:IonizationReaction2} becomes complete and no interstitial lithium is left over.

\begin{figure}[htbp]
\centering
  \includegraphics[width=120mm]{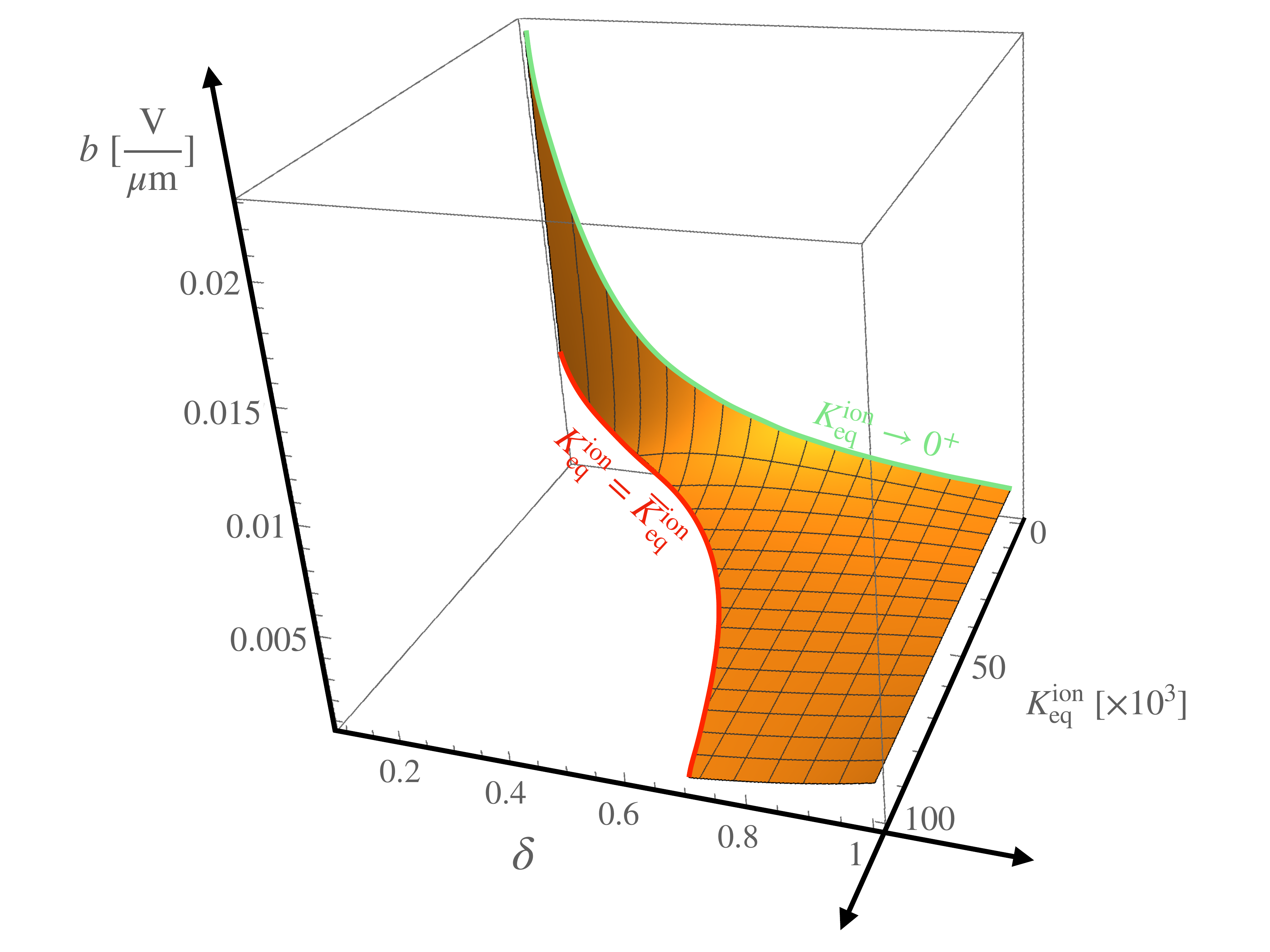}
\caption{ \em Steady state representation of coefficient $b$ (i.e. the gradient of the electric potential) as a function of $\delta$ and $K_{\rm eq}^{\rm ion}$. With the material properties of table \ref{Tab:Input} , ${\underline K}_{\rm eq}^{\rm ion} = 0$. }
\label{fig:phi_draw}
\end{figure}

As per eq. \eqref{eq:steadystatephi}, the electric potential is linear. Coefficient $b$ can be derived from eq. \eqref{eq:Nernst-Planck_ss} as
\begin{equation}
\label{eq:b_ss}
b = 
-
    \frac{I(t)}{A} \, \frac{RT}{F^2} 
    \;
    \frac{1}{\diffusivity_{{\rm Li}^+} \, c^{\rm ss}_{{\rm Li}^+}
 +
 \diffusivity_{{\rm Li}^+_{\rm hop}} \, c^{\rm ss}_{{\rm Li}^+_{\rm hop}}
} 
 \; ,
\end{equation}
under the condition that the denominator is positive, i.e.
\begin{align}
\label{eq:bcond_ss}
 \, K_{\rm eq}^{\rm ion} 
 \, 
>  \; {\overline K}_{\rm eq}^{\rm ion}  \;
 \frac{\diffusivity_{{\rm Li}^+_{\rm hop}} } {  \diffusivity_{{\rm Li}^+_{\rm hop}} - \diffusivity_{{\rm Li}^+} }
 \; ,
\end{align}
which in turn puts a condition on diffusivities in order to achieve a steady state condition, i.e.
\begin{align}
\label{eq:lowerbound_ss}
 \diffusivity_{{\rm Li}^+_{\rm hop}} < \diffusivity_{{\rm Li}^+}
 \; 
\end{align}
and sets  $ {\underline K}_{\rm eq}^{\rm ion} =0 $ as a {\em{lower bound}} on $ K_{\rm eq}^{\rm ion} $.
%
%
The values taken by $b$ are plot in fig. \ref{fig:phi_draw} as a function of $\delta$ and $K_{\rm eq}^{\rm ion}$.

\begin{figure}[!htbp]
\centering
\begin{subfigure} {0.495\textwidth}
  \includegraphics[width=80mm]{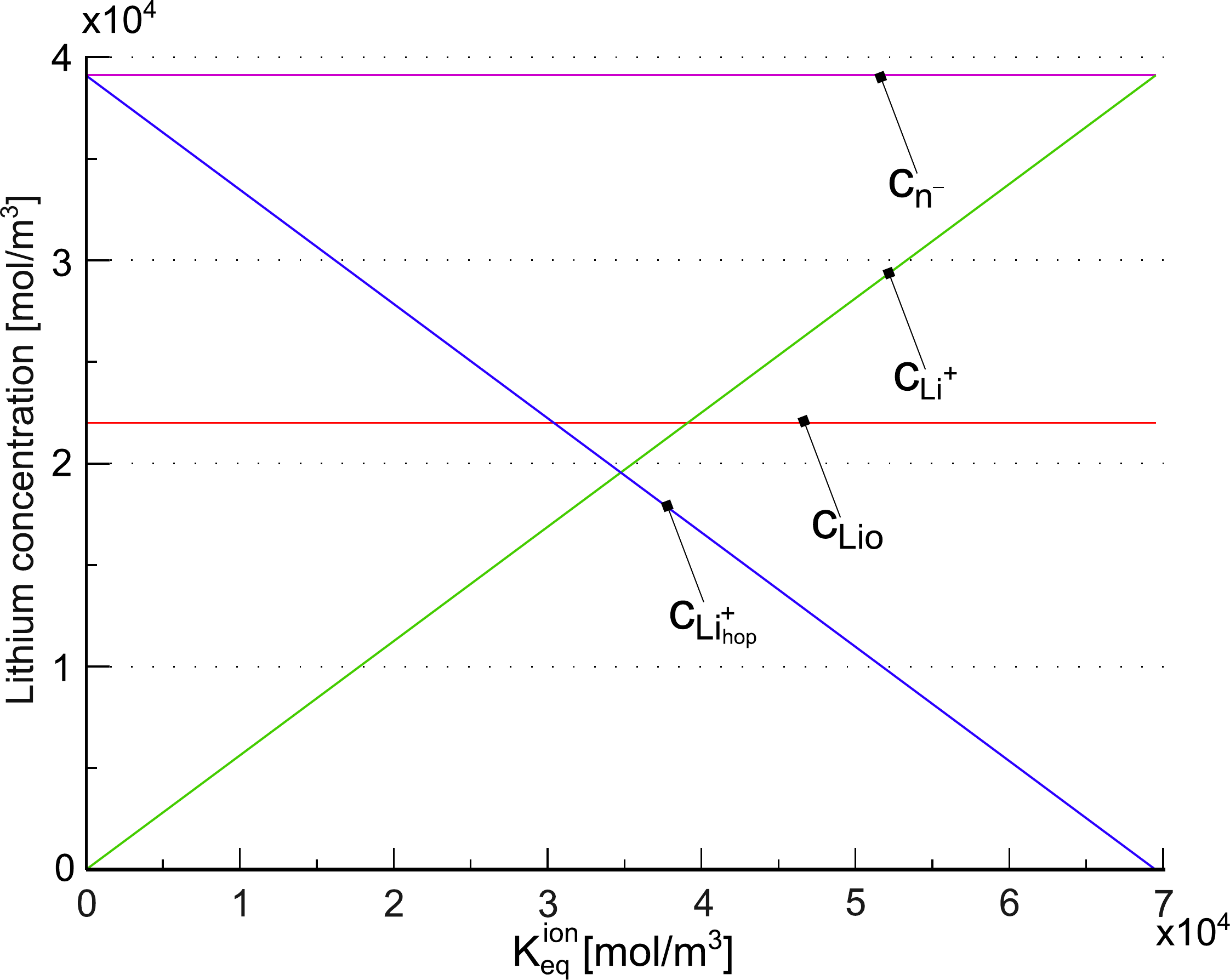}
\caption{  }
\label{subfig:Steady}
\end{subfigure}
\begin{subfigure} {0.495\textwidth}
  \includegraphics[width=80mm]{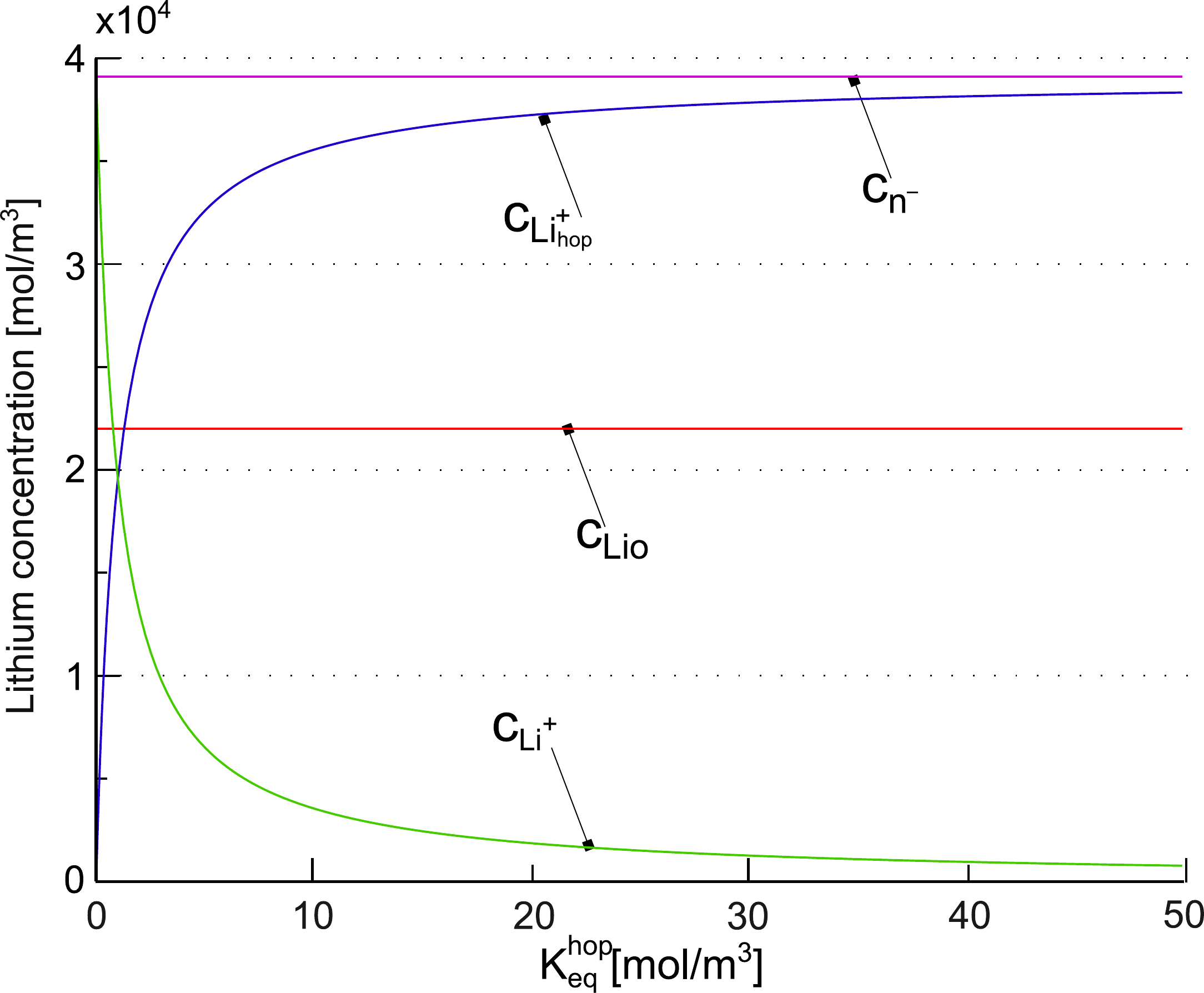}
\caption{  }
\label{subfig:Steady_1}
\end{subfigure}
\caption{ \it Concentration of species  $c_{\rm Li_0}^{\rm ss}$, $c_{\rm n^-}^{\rm ss}$, $c_{{\rm Li}^+}^{\rm ss}$, $c_{{\rm Li}^+_{\rm hop}}^{\rm ss}$ as a function of $K_{\rm eq}^{\rm ion}$ (a) and $K_{\rm eq}^{\rm hop}$ (b) at $\delta = 0.64$ and $c_0 = 61141$.}
\label{fig:Steady}
\end{figure}

\begin{figure}[!htbp]
\centering
\begin{subfigure} {0.495\textwidth}
  \includegraphics[width=80mm]{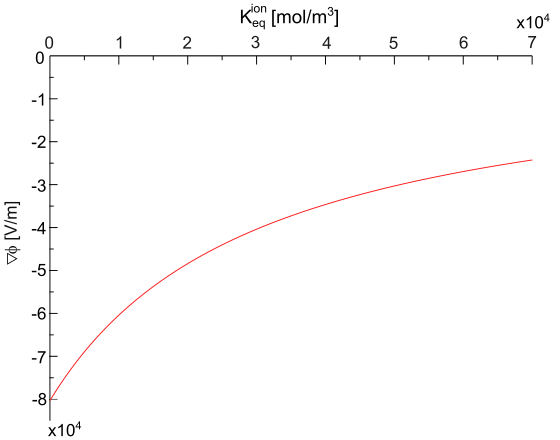}
\caption{  }
\label{subfig:Steady1}
\end{subfigure}
\begin{subfigure} {0.495\textwidth}
  \includegraphics[width=80mm]{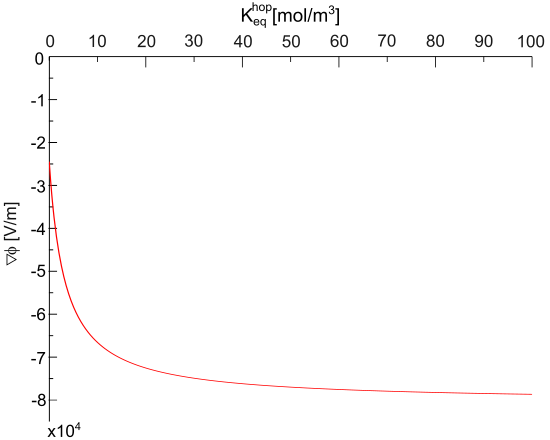}
\caption{  }
\label{subfig:Steady1_1}
\end{subfigure}
\caption{ \it Electric potential as a function of $K_{\rm eq}^{\rm ion}$ (a) and  $K_{\rm eq}^{\rm hop}$ (b). The asymptotic values of the gradient of the electric potential are $b=-I(t)RT/(\delta A F^2\diffusivity_{{\rm Li}^+_{\rm hop}}c_0)$ for $K_{\rm eq}^{\rm ion}=0$ and $b=-I(t)RT/(\delta AF^2\diffusivity_{{\rm Li}^+}c_0)$ for $K_{\rm eq}^{\rm ion}={\overline K}_{\rm eq}^{\rm ion}$.}
\label{fig:Steady1}
\end{figure}

\bigskip
The influence of the equilibrium constants $K_{\rm eq}^{\rm ion}$ and $K_{\rm eq}^{\rm hop}$ on the concentration of the species $c_{\rm Li_0}^{\rm ss}$, $c_{\rm n^-}^{\rm ss}$, $c_{{\rm Li}^+}^{\rm ss}$, $c_{{\rm Li}^+_{\rm hop}}^{\rm ss}$ and on the gradient of the electric potential are shown in Figures \ref{fig:Steady} and \ref{fig:Steady1} for the particular value $\delta=0.64$.

\subsection{Single discharge response of the electrolyte}
\label{subsec:transient}

The steady state solution \eqref{eq:ion_react_equilibrium_5}, \eqref{eq:steadystatedelta} and \eqref{eq:b_ss} turns out to be the numerically simulated response of the electrolyte when initial conditions are imposed to be at equilibrium, according to eqs. \eqref{eq:ion_react_equilibrium_2}-\eqref{eq:ion_react_equilibrium_4}, and boundary conditions on fluxes \eqref{eq:MassFluxBC} are chosen as such as to maintain such steady state solution. In the battery operation, though, boundary conditions may deviate from this ideal state and transient behaviors have been observed in a companion paper \cite{CabrasEtAl2021b}. For this sake, time dependent solutions are here sought for, by purposely altering the initial conditions. The response of the system is studied at different values of material parameters.

\subsubsection{Case study 1 : $K_{\rm eq}^{\rm ion} = {\overline K}_{\rm eq}^{\rm ion} $} 
The two reaction rate constants that describe the ionization reaction \eqref{eq:IonizationReaction1}, i.e the lithium ion recombination rate $k_b^{\rm ion}$ and the generation rate $k_f^{\rm ion}$, were identified in \cite{RaijmakersEtAlEA2020}, $k_b^{\rm ion} =\!8.00\cdot10^{-7}$ $\rm m^3mol^{-1}s^{-1}$ and $k_f^{\rm ion}\!=\!5.56\cdot10^{-2}$ $\rm s^{-1}$. Their ratio is $K_{\rm eq}^{\rm ion} = {\overline K}_{\rm eq}^{\rm ion} =69518 \; \rm mol \, m^{-3}$. As stated in section \ref{subsec:chem_kin}, in such a case $K_{\rm eq}^{\rm hop} =0$ and if at initial time $c_{{\rm Li}^+_{\rm hop}} = 0$ then no hopping lithium is further generated and no hopping takes place, with steady state charge transport of pure interstitial type. 
For the sake of understanding the transient response of the electrolyte, analyses have been carried out perturbing the initial ionic concentration from the state of equilibrium, as follows:
\begin{equation}
c_{{\rm Li_0}}=(1-\delta)c_0; \qquad c_{{\rm n}^-}=\delta c_0; \qquad c_{{\rm Li}^+}=0.9\cdot c_{{\rm n}^-};\qquad c_{{\rm Li}^+_{\rm hop}}=0.1\cdot c_{{\rm n}^-}  \qquad {\vect{x}} \in V, \; t=0 \; .
\label{eq:initcondpb1}
\end{equation}
Since an initial concentration of hopping lithium is now available, it will be transformed in interstitial by reaction \eqref{eq:IonizationReaction2}, with pace ruled by the constant $k_b^{\rm hop}$, with zero forward rate by case-study hypothesis. 

\begin{figure}[!htbp]
\includegraphics[width=16cm]{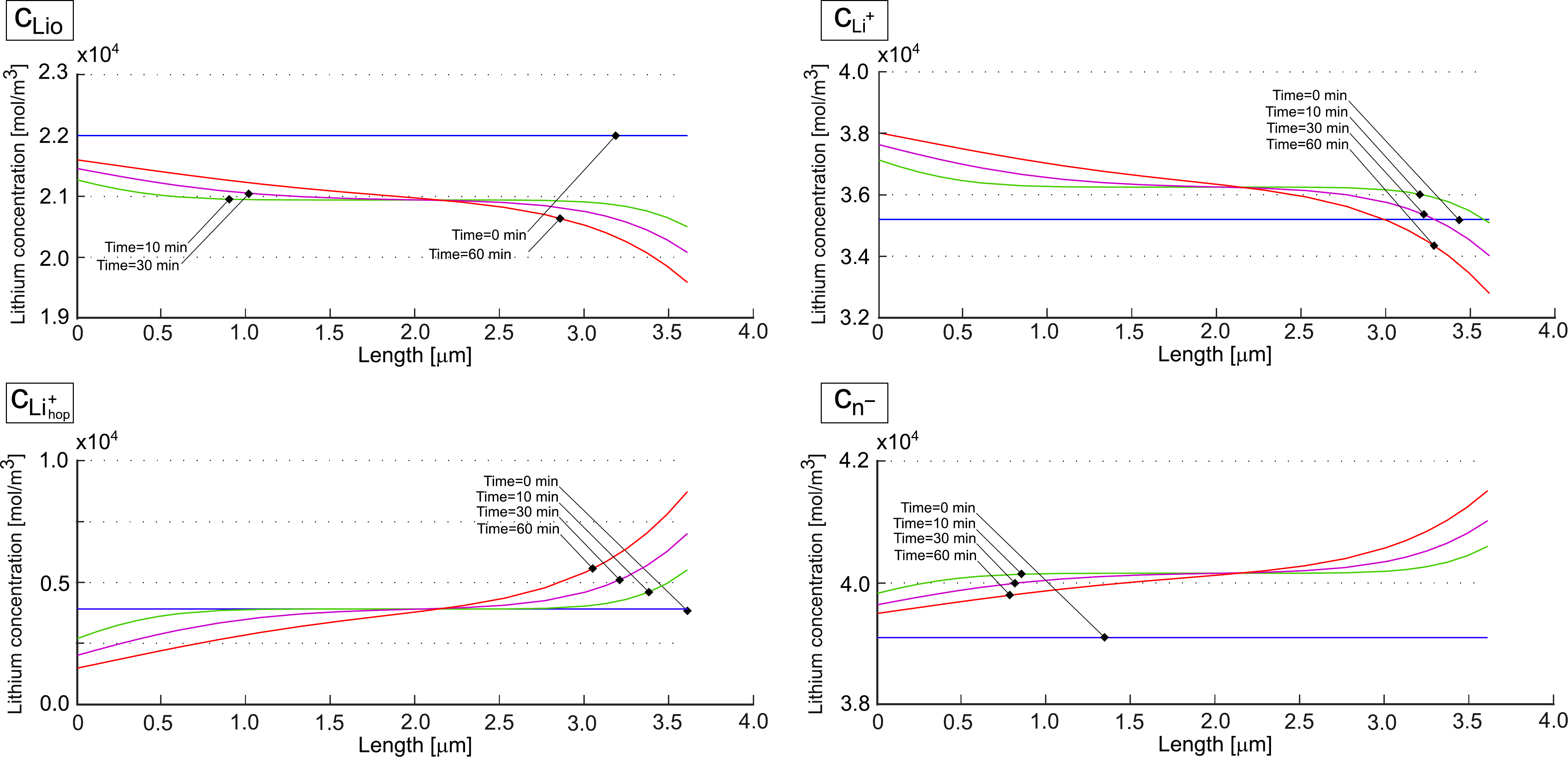}
\caption{\it \it Concentrations in the electrolyte. Four instants are  considered: besides the initial time, 10 minutes, 30 minutes, 1 hour.}
\label{fig:ConcentrationTransient2}
\end{figure}

\begin{figure}[!htbp]
\centering
\begin{subfigure} {0.495\textwidth}
  \includegraphics[width=80mm]{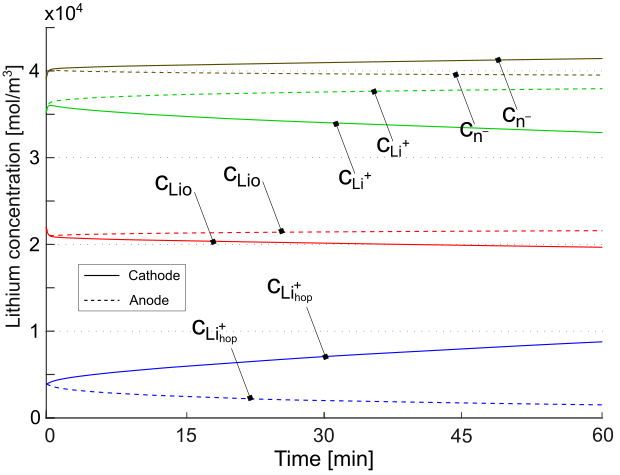}
\caption{  }
\label{subfig:ConcentrationAnodeCathodeforTransient2}
\end{subfigure}
\begin{subfigure} {0.495\textwidth}
  \includegraphics[width=80mm]{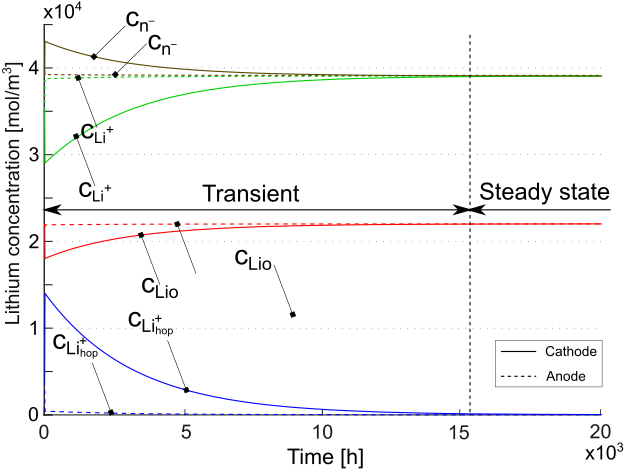}
\caption{  }
\label{subfig:ConcentrationAnodeCathodeforTransientZoom2}
\end{subfigure}
\caption{\it Ionic concentration profile at anode and cathode for the different species. The dashed lines represent the values at the anode, the continuous lines at the cathode. a) Zoom around $t=0$ to show how concentration depart from initial values for the transient behavior. b) A complete time-span evolution, showing how the steady asymptotic behavior is  recovered.}
\label{fig:ConcAnoCatTransient2}
\end{figure}

\begin{figure}[!htbp]
\centering
\includegraphics[width=160mm]{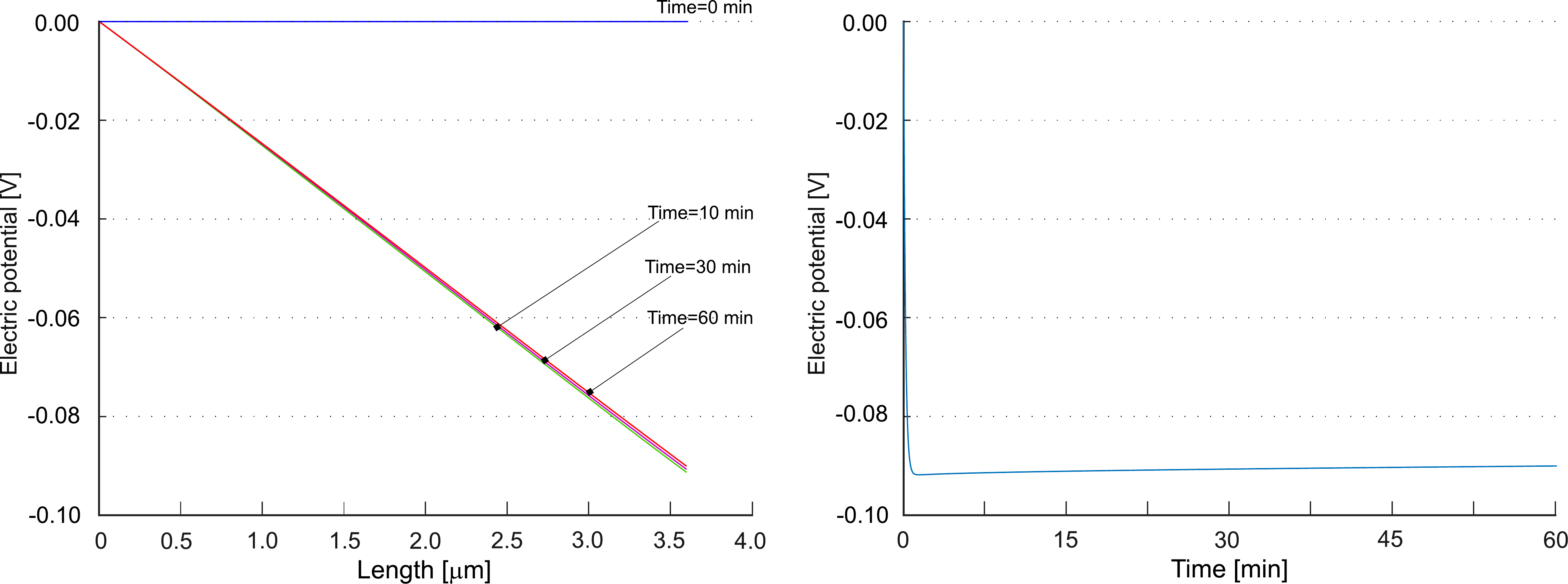}
\caption{\it a)The electric potential $\phi(x)$, parametrized in time. b) Its value $\phi(L)$ at the cathode interface.}
\label{fig:ElPotentialTransient2}
\end{figure}

\bigskip
{\em{Consider first $k_b^{\rm hop}=k_b^{\rm ion}$}}, to which figures \ref{fig:ConcentrationTransient2}-\ref{fig:ElPotentialTransient2} pertain. 
The transient evolution is clearly visible for species concentrations profiles at different instants (10, 30, 60 minutes, respectively) in fig. \ref{fig:ConcentrationTransient2}. The oxygen-bound lithium concentration $c_{{\rm Li_0}}$ decreases in the whole electrolyte in the transient period, whereas vacancies do the opposite. Figure \ref{fig:ConcAnoCatTransient2}, which focuses on the species concentrations at anode and cathode, shows that the concentration profiles acquire a steady state regime only after a very long time for the parameters at hand: actually the time frame is so large to be unrealistically overstated. One may argue therefore that, for some selections of parameters, the transient regime might drive the whole charging/discharging evolution. The time required to approach a steady state regime is governed by the values of the reaction rate constants.
Dotted lines in Fig. \ref{fig:ConcAnoCatTransient2} denote the concentration of species at the interface with the anode, whereas continuous lines are used for the cathode interface. 


Figure \ref{fig:ElPotentialTransient2}a depicts the evolution of the electric potential $\phi(x)$ along the solid electrolyte at 10, 30, 60 minutes. It shows a tendency to reach the steady state much faster than concentrations. Figure \ref{fig:ElPotentialTransient2}b reports the evolution of the electric potential at the cathode interface.

The deviation from perfect electroneutrality condition is estimated by the ratio $\rho_{el}$
\begin{equation*}
\rho_{el}(t) = \sup_{0 \le x \le L} \;
\frac{ c_{\rm Li^+}+c_{{\rm Li}^+_{\rm hop}} - c_{\rm n^-} } {c_{\rm Li^+}+c_{{\rm Li}^+_{\rm hop}}+ c_{\rm n^-} }
\; .
\end{equation*}
%
Electroneutrality is well approximated by the numerical solution, since $\rho_{el}(t) \sim 10^{-7}$ during the time-span of the discharge process. 

%

\begin{figure}[!htb]
\includegraphics[width=16cm]{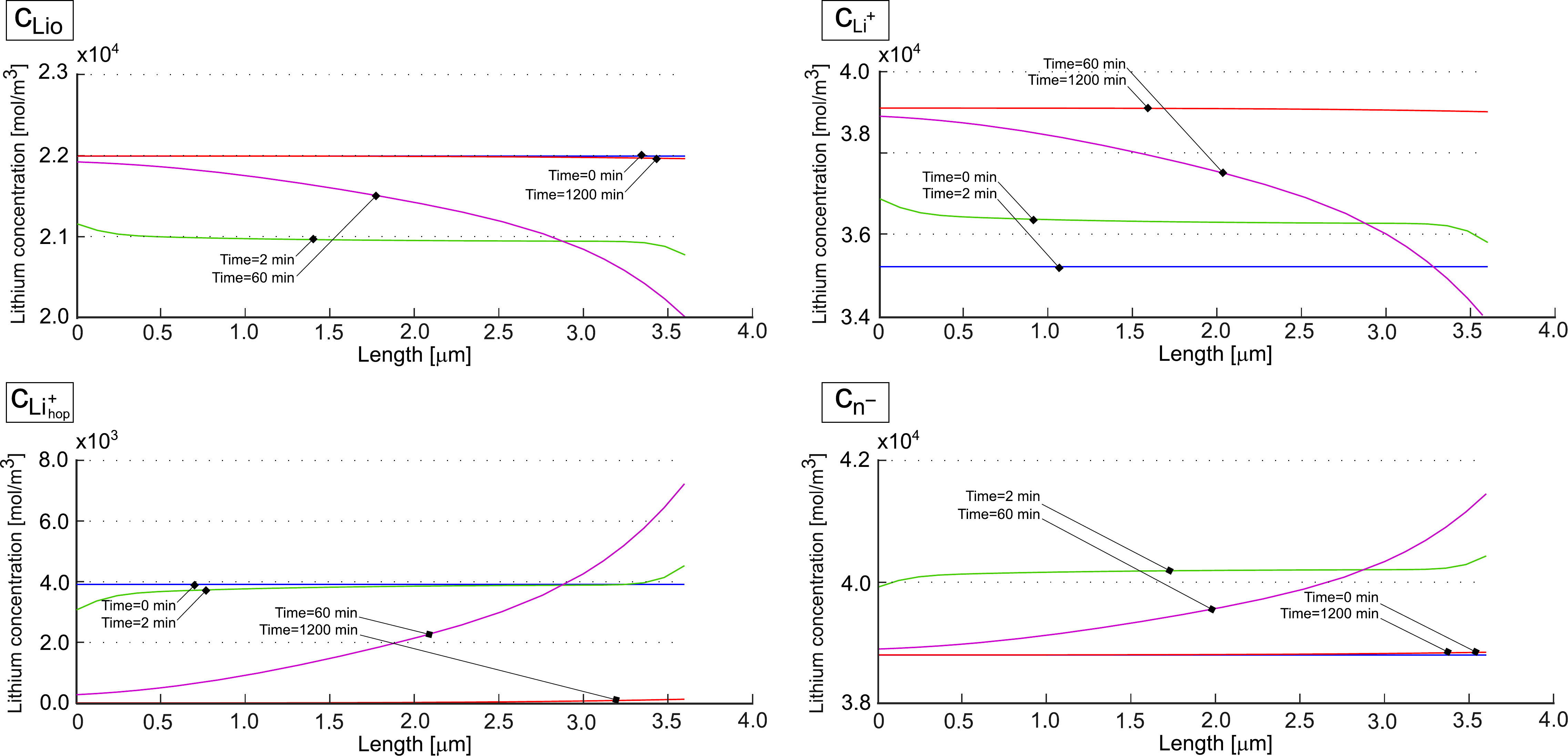}
\caption{\it Concentrations in the electrolyte. Four instants are  considered, initial one (blue), 1 hour (red), 10 minutes (green) and 30 minutes (yellow). $k_b^{\rm ion} =\!8.00\cdot10^{-7}$, $k_f^{\rm ion}\!=\!5.56\cdot10^{-2}$, $k_b^{\rm hop}=1000k_b^{\rm ion}$ and $k_f^{\rm hop}=0$.}
\label{fig:ConcentrationTransient3}
\end{figure}

\begin{figure}[!htbp]
\centering
\includegraphics[width=8cm]{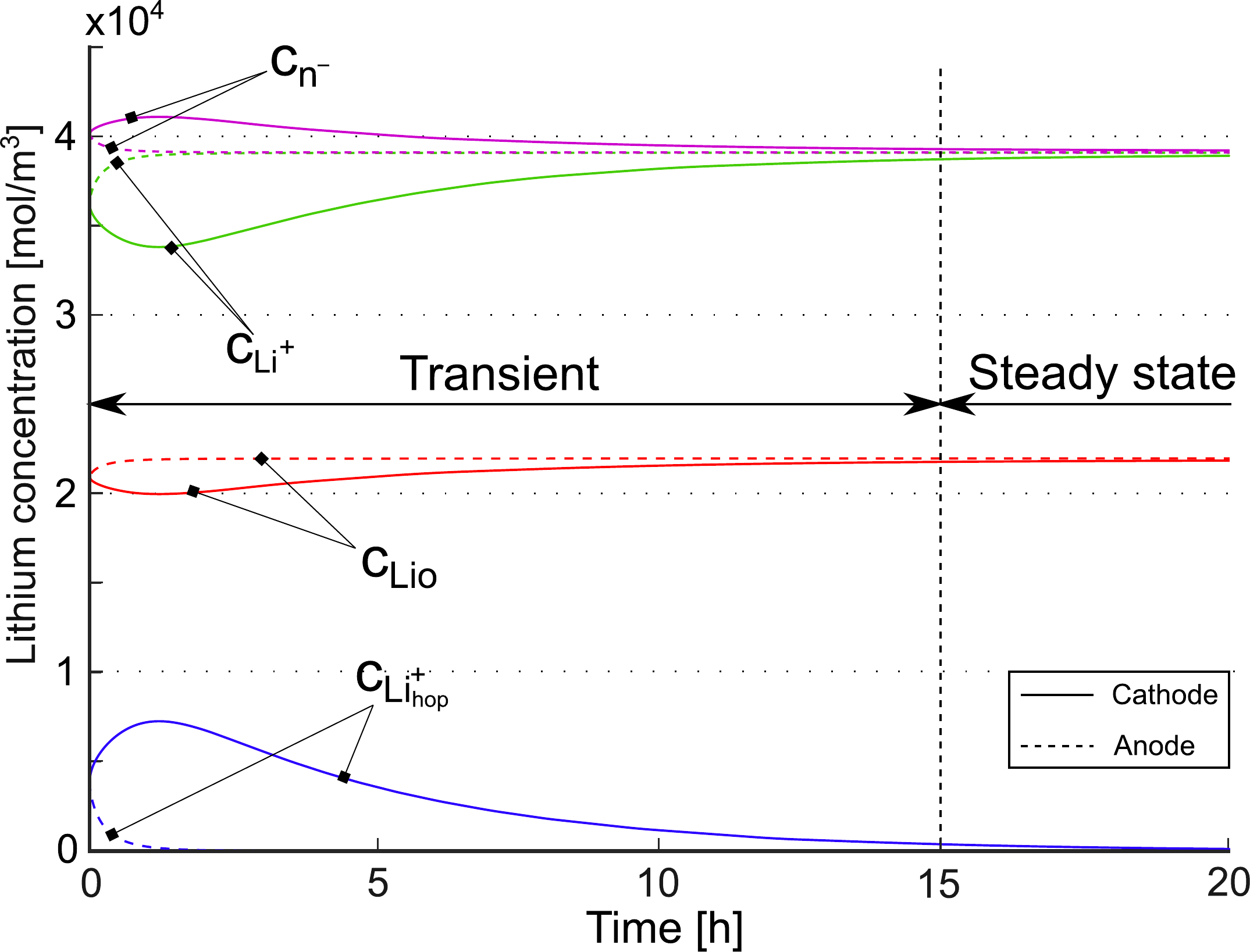}
\caption{\it Ionic concentration profile at the interface with the anode and the cathode. The dashed lines represent the values in the anode, the continuous lines in the cathode. $k_b^{\rm ion} =\!8.00\cdot10^{-7}$, $k_f^{\rm ion}\!=\!5.56\cdot10^{-2}$, $k_b^{\rm hop}=1000k_b^{\rm ion}$ and $k_f^{\rm hop}=0$.}
\label{fig:ConcAnoCatTransient3}
\end{figure}

\bigskip
To highlight the influence of $k_b^{\rm hop}$, 
{\em{we took $k_b^{\rm hop}=1000k_b^{\rm ion}$}} while keeping all other parameters unaltered. The increment of $k_b^{\rm hop}$ reduces significantly the time to reach the steady state regime, since the undesired initial hopping lithium is much quickly converted into interstitial. Compare in this regard Fig.\ref{fig:ConcAnoCatTransient3} with Fig.\ref{fig:ConcAnoCatTransient2}-b and Fig. \ref{fig:ConcentrationTransient3}a with  Fig. \ref{fig:ConcentrationTransient2}.


\subsubsection{Case study 2 : $K_{\rm eq}^{\rm ion} = 0.85 \, {\overline K}_{\rm eq}^{\rm ion} $}

In a second case study, the equilibrium constant $K_{\rm eq}^{\rm ion}$ has been reduced by fifteen percent, i.e. $K_{\rm eq}^{\rm ion}=59090$ ${\rm mol / m^3}$, while the rate constant $k_b^{\rm ion}=\!8.00\cdot10^{-7}$ $\rm m^3mol^{-1}s^{-1}$ as for the previous case. The equilibrium constant for reaction \eqref{eq:IonizationReaction2} becomes $K_{\rm eq}^{\rm hop}=0.1765$, from eq.\eqref{eq:keq_and_delta}, and as first we consider $k_b^{\rm hop}=k_b^{\rm ion}$. Initial conditions are taken as in \eqref{eq:initcondpb1}.

\begin{figure}[!htb]
\includegraphics[width=16cm]{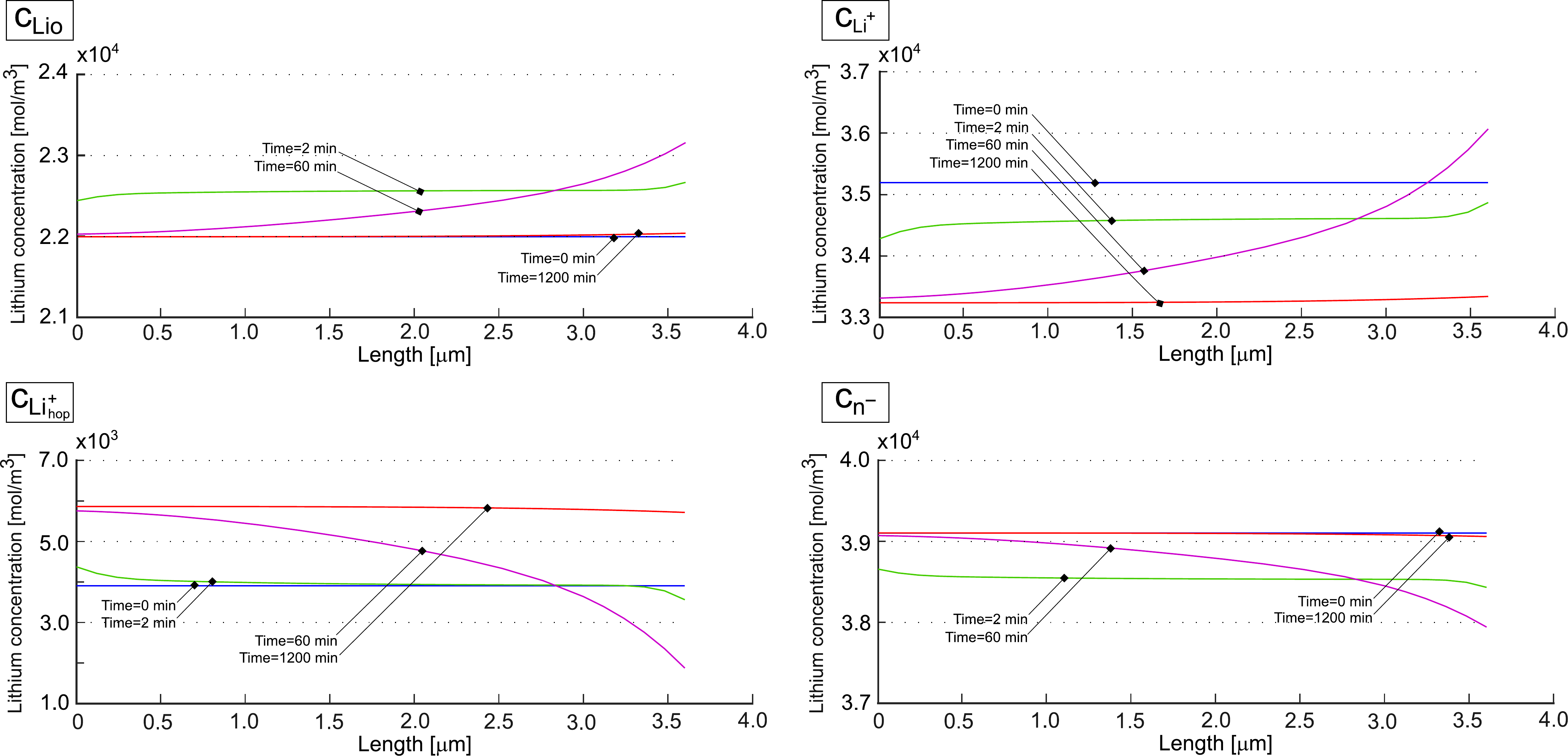}
\caption{\it Concentrations in the electrolyte. Four instants are  considered, initial one (blue), 1 hour (red), 10 minutes (green) and 30 minutes (yellow). $k_b^{\rm ion} =\!8.00\cdot10^{-7}$, $k_f^{\rm ion}\!=\!4.726\cdot10^{-2}$, $k_b^{\rm hop}=1000k_b^{\rm ion}$ and $k_f^{\rm hop}=1.4118\cdot10^{-4}$.}
\label{fig:ConcentrationTransient3b}
\end{figure}

\begin{figure}[!htbp]
\centering
\includegraphics[width=80mm]{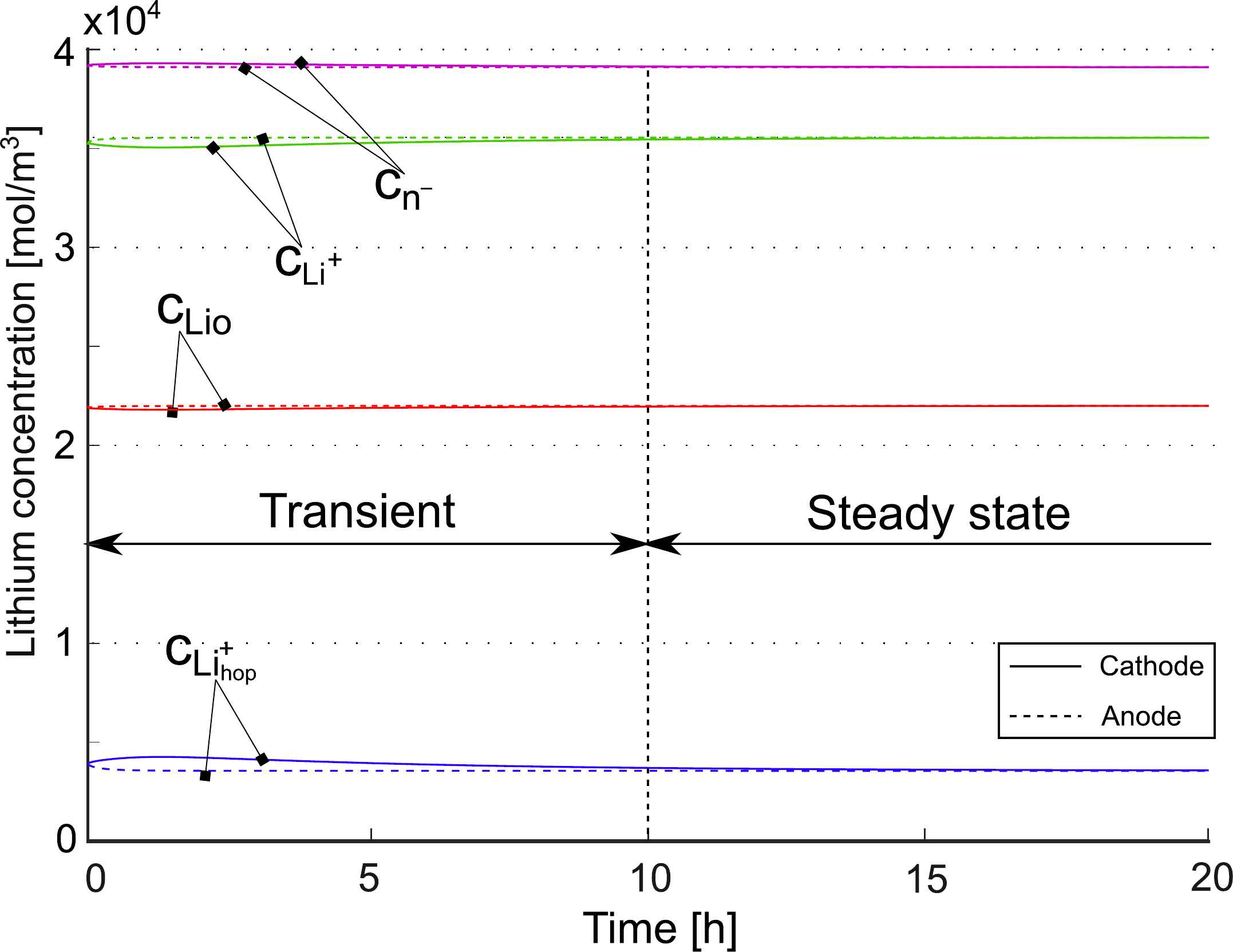}
\caption{\it Ionic concentration profile at the interface wit the anode and the cathode. The dashed lines represent the values in the anode, the continuous lines in the cathode. $k_b^{\rm ion} =\!8.00\cdot10^{-7}$, $k_f^{\rm ion}\!=\!4.726\cdot10^{-2}$, $k_b^{\rm hop}=1000k_b^{\rm ion}$ and $k_f^{\rm hop}=1.4118\cdot10^{-4}$.}
\label{fig:ConcAnoCatTransient3b}
\end{figure}

The concentration of the different species in the electrolyte is plotted in Fig.\ref{fig:ConcentrationTransient3b}. The overall response is similar to what discussed in fig. \ref{fig:ConcentrationTransient2}, but the steady state is reached more rapidly as the curves at 20 h display. Similar conclusion can be inferred from Fig.\ref{fig:ConcAnoCatTransient3b}, where the concentrations at the two electrodes are shown.  
%
%

As already noticed in the former case-study, 
the evolution of the electric potential $\phi(x)$ along the solid electrolyte shows a tendency to reach the steady state much faster than concentrations.
\begin{figure}[!htbp]
\centering
\includegraphics[width=8cm]{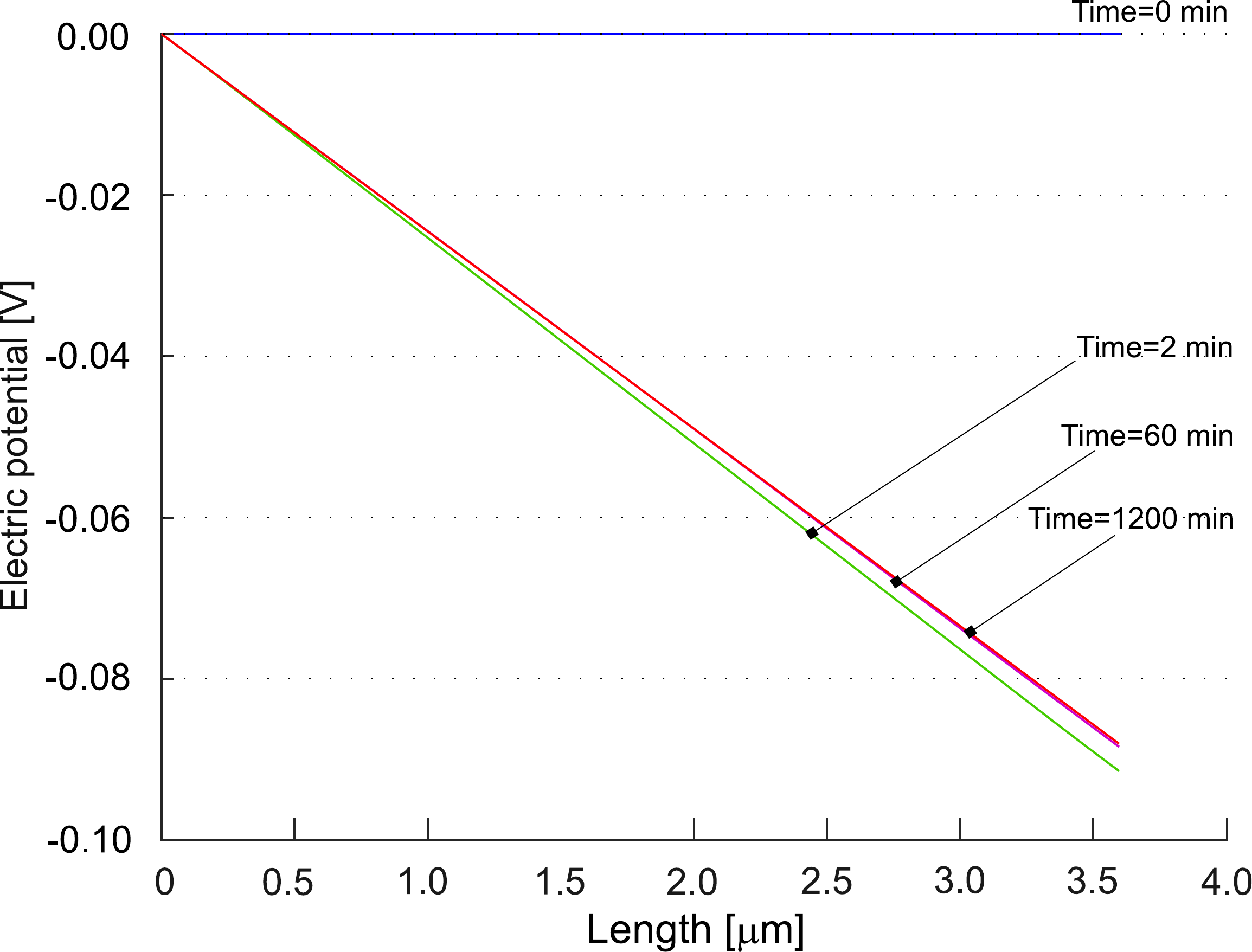}
\includegraphics[width=8cm]{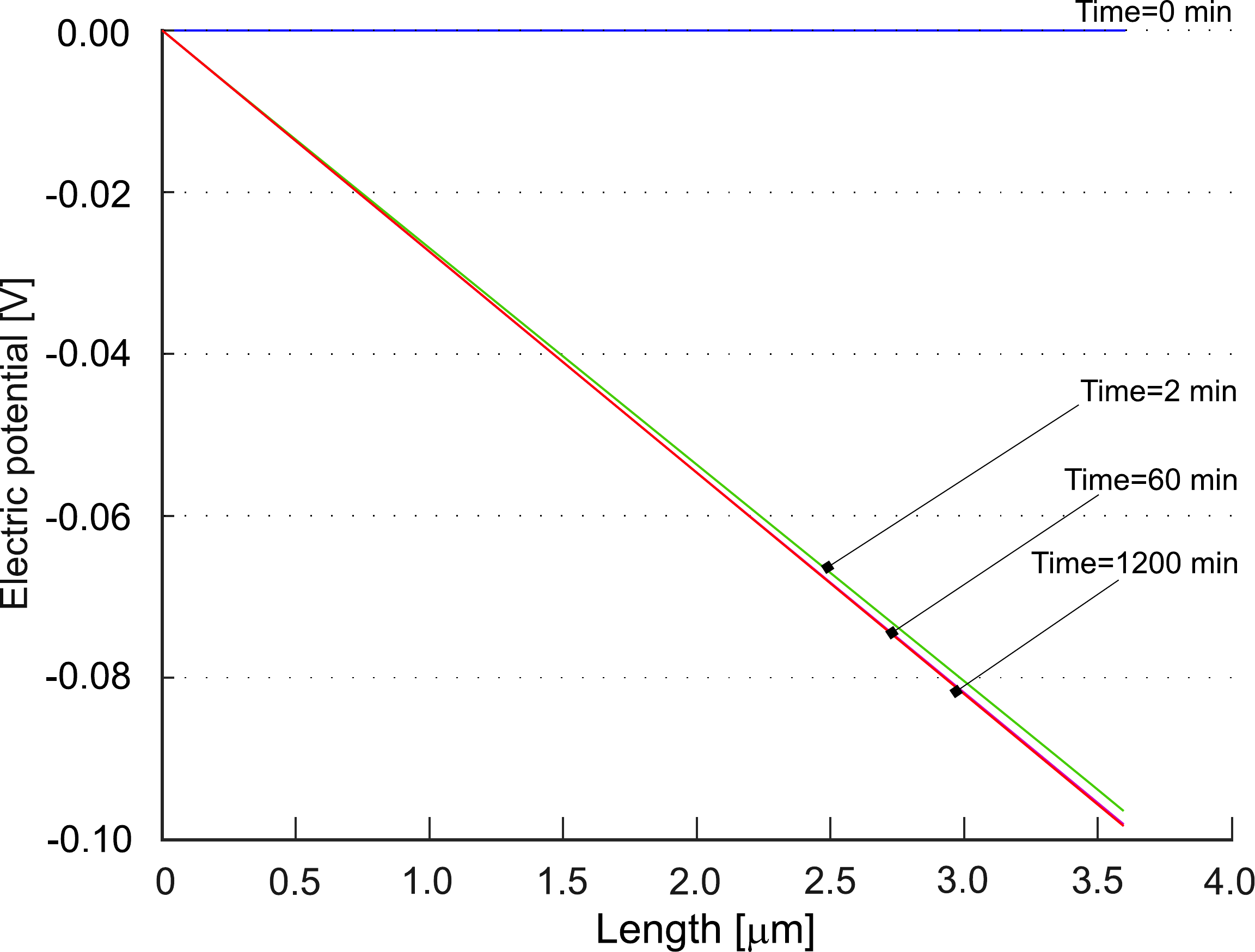}
\caption{\it The electric potential $\phi(x)$ inside the electrolyte. a) $k_b^{\rm ion} =\!8.00\cdot10^{-7}$, $k_f^{\rm ion}\!=\!5.56\cdot10^{-2}$, $k_b^{\rm hop}=1000k_b^{\rm ion}$ and $k_f^{\rm hop}=0$. b) $k_b^{\rm ion} =\!8.00\cdot10^{-7}$, $k_f^{\rm ion}\!=\!4.726\cdot10^{-2}$, $k_b^{\rm hop}=1000k_b^{\rm ion}$ and $k_f^{\rm hop}=1.4118\cdot10^{-4}$.}
\label{fig:ElPotentialTransient3}
\end{figure}

\section{Sensitivity analysis of the model parameters}
\label{sec:sensitivity}

In this section we perform a SA to our model, in order to identify the importance of each parameter and its contribution to the variability of the model predictions derived in section \ref{subsec:Steady_state_response} at {\em{steady state}}.  In ideal scenario, all the model parameters should be estimated as accurately as possible from carefully designed physical experiments. The cost  and time restrictions, however, can limit the access to experimental data required for model calibration and validation. Conducting a SA prior to the physical experiments can help in identifying the {\it essential} parameters to be estimated. The {\it non-essential} parameters, on the other hand, can be set to nominal values obtained from the literature or any prior physical knowledge. The essential parameters can be defined as those with high {\em{sensitivity index}}. 
Simulations of the real physical setting can be used to identify essential parameters by varying each parameter within a given range and properly scrutinizing the output. 

Specifically, since eq. \eqref{eq:Khop_identity} holds, three parameters are required to define the steady state solution, i.e. the maximal concentration of host-sites $c_0$, the fraction of Li that resides in equilibrium in the mobile state, $\delta$,  and the equilibrium constant of reaction \eqref{eq:IonizationReaction1}, $K_{\rm eq}^{\rm ion}$.
%
%
Assuming that $c_0$ can be estimated with high accuracy on theoretical grounds, the interest is to study the effect of $\delta$ and $K_{\rm eq}^{\rm ion}$ on the steady state solutions \eqref{eq:steadystatedelta} and \eqref{eq:b_ss}.

The Sobol' index, which is a variance-based method, is used to this aim. In the probabilistic setting, the model parameters are assumed to be random variables and a surrogate model is built to map the inputs to the corresponding output. Once the surrogate is constructed, Sobol decomposition provides the sensitivity indices \cite{saltelli2004sensitivity}. We use the Bayesian Hybrid Modeling (GEBHM) approach \cite{ghosh2020advances, zhang2020remarks}, 
a probabilistic machine learning method that enables SA, calibration, multi-fidelity modeling and uncertainty quantification.   

\begin{figure} [t]
\centering
\begin{subfigure} {0.27\textwidth}
  \includegraphics[width=\textwidth]{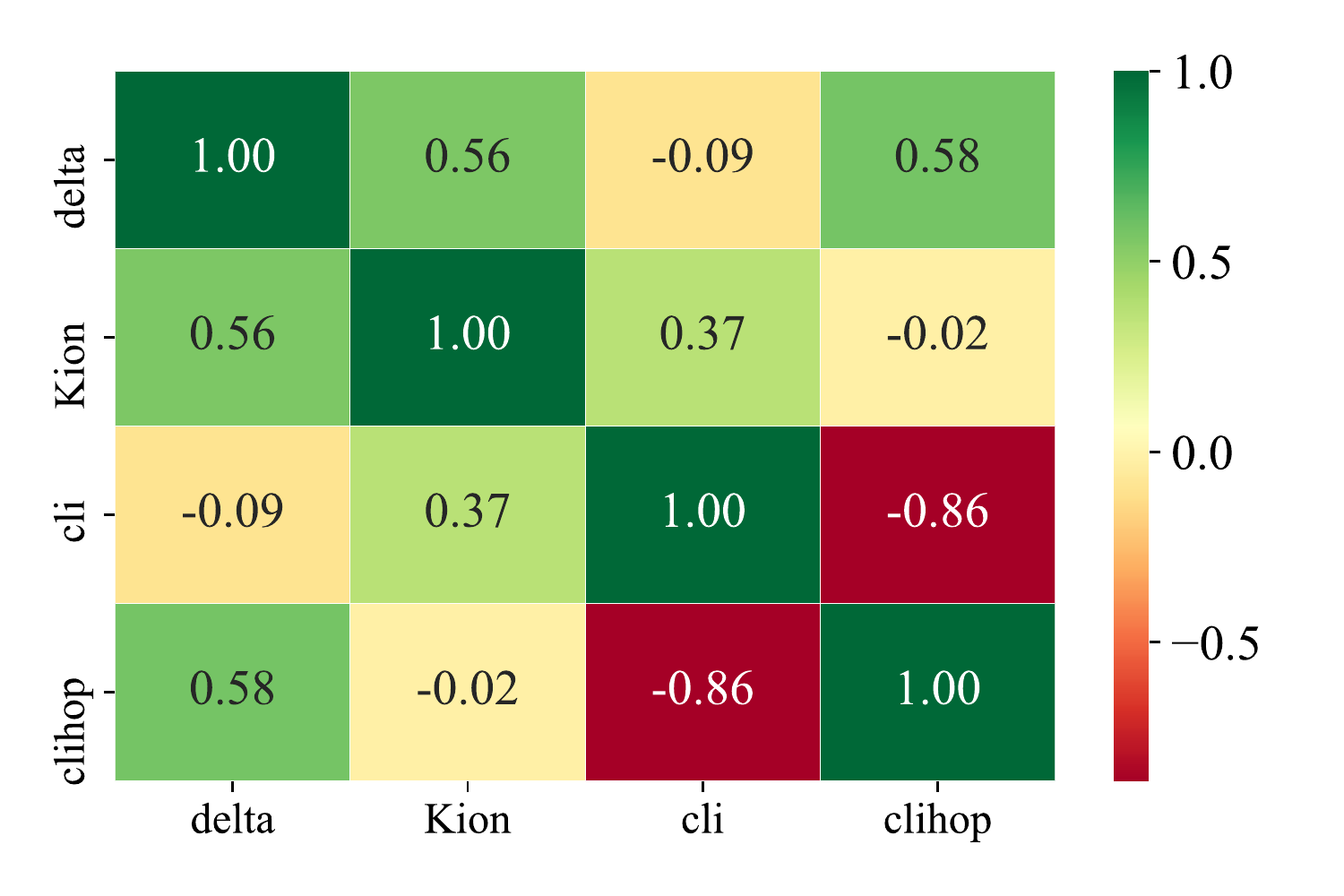} 
\caption{}
\label{subfig:case_a_corr}
\end{subfigure}
\begin{subfigure} {0.34\textwidth}
  \includegraphics[width=\textwidth]{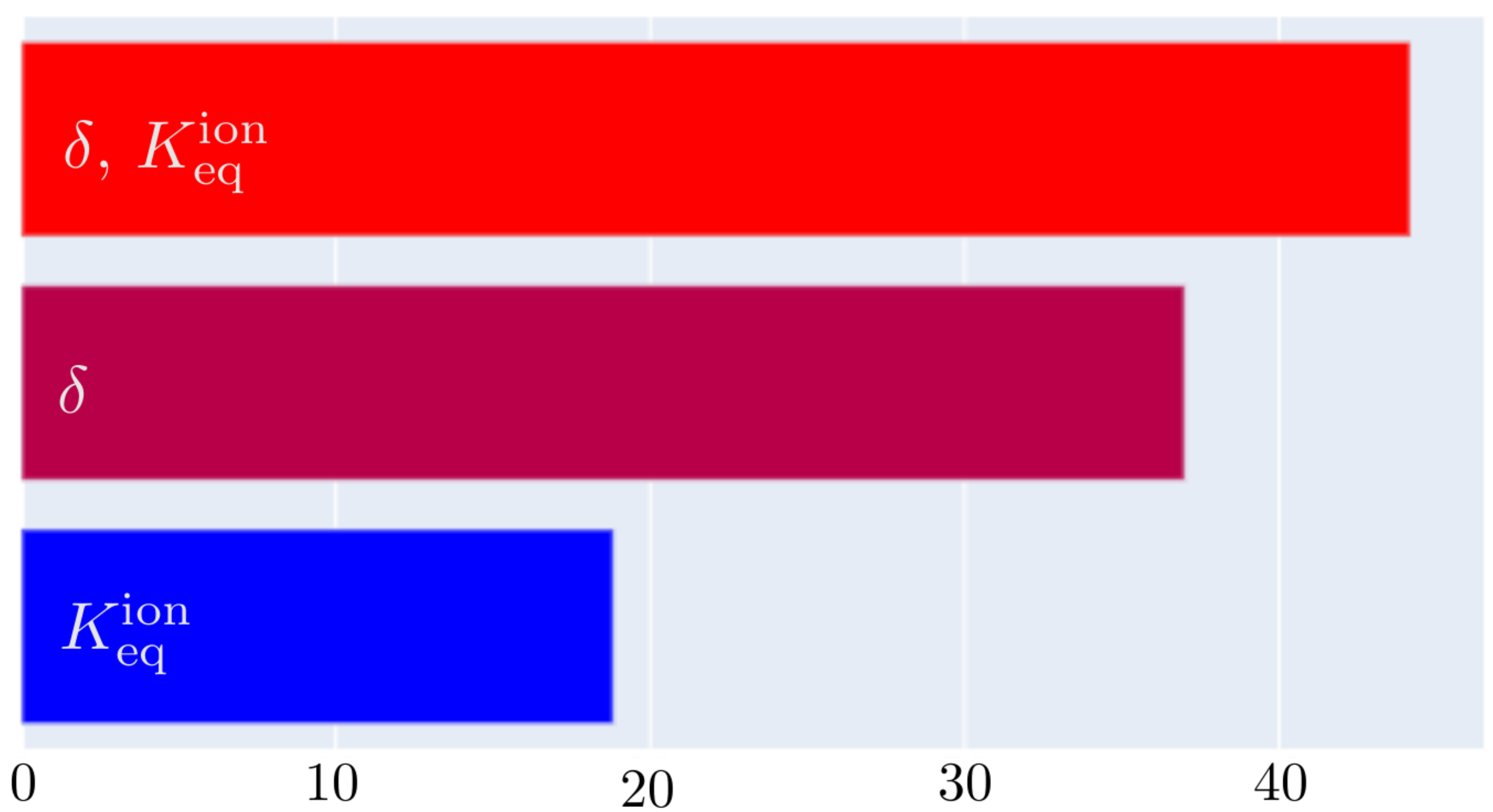}
\caption{$c_{{\rm Li}^+}^{\rm ss} $ }
\label{subfig:barplotcliss}
\end{subfigure}
\begin{subfigure} {0.35\textwidth}
  \includegraphics[width=\textwidth]{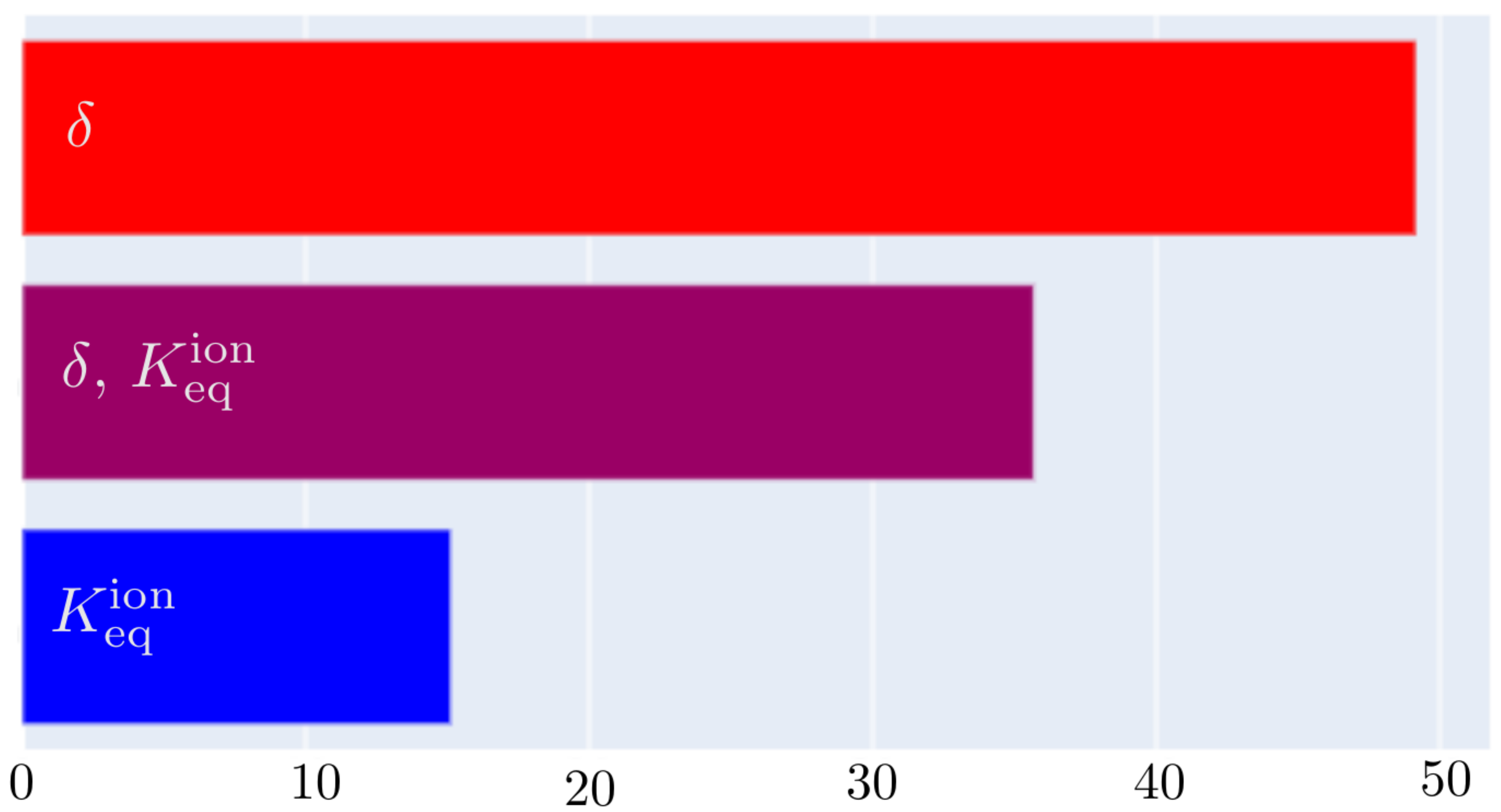}
\caption{$ c_{{\rm Li}^+_{\rm hop}}^{\rm ss}$}
\label{subfig:barplotclisshop}
\end{subfigure}
%
 \caption{\em{The correlation matrix  between the input and output (a). The total Sobol indices for: (b) $c_{{\rm Li}^+}^{\rm ss} $  and (c) $ c_{{\rm Li}^+_{\rm hop}}^{\rm ss}$ } }
 \label{bar_plot}
 \end{figure}

The matrix in Fig.~\ref{subfig:case_a_corr} shows the correlation between the input parameters  $\delta$ and $K_{\rm eq}^{\rm ion}$ and the output uniform concentrations $c_{{\rm Li}^+}^{\rm ss} $ and $ c_{{\rm Li}^+_{\rm hop}}^{\rm ss}$. We notice that $\delta$ has a small negative correlation with $c_{{\rm Li}^+}^{\rm ss} $ and relatively large positive correlation with $ c_{{\rm Li}^+_{\rm hop}}^{\rm ss}$. On the other hand,  $K_{\rm eq}^{\rm ion}$ has a small negative correlation with $ c_{{\rm Li}^+_{\rm hop}}^{\rm ss}$ and positive correlation with $c_{{\rm Li}^+}^{\rm ss} $. This is expected given the structure of the expression relating the inputs to outputs (see Eq. \eqref{eq:steadystatedelta}). 
Fig.~\ref{subfig:barplotcliss} shows  Sobol indices for the input parameters  $\delta$ and $K_{\rm eq}^{\rm ion}$ representing the percentage of their contribution to the total variance of the outputs $c_{{\rm Li}^+}^{\rm ss} $ and $ c_{{\rm Li}^+_{\rm hop}}^{\rm ss}$, respectively. The values of the total Sobol indices can be viewed as an indicator of the relevance of each parameter. For example in  Fig.~\ref{subfig:barplotcliss},  $\delta$ contributes around $37\%$ to the variability of $c_{{\rm Li}^+}^{\rm ss} $, whereas the interaction between $\delta$ and $K_{\rm eq}^{\rm ion}$ contributes around $48\%$. Similarly, Fig.~\ref{subfig:barplotclisshop} shows that  $\delta$ contributes around $49\%$ to the  variability of $ c_{{\rm Li}^+_{\rm hop}}^{\rm ss}$. These results show that for the concentration $ c_{{\rm Li}^+_{\rm hop}}^{\rm ss}$, $\delta$  is the most influential  parameter, while the interaction  between  $\delta$ and $K_{\rm eq}^{\rm ion}$ is the most influential on the variability if the output f $c_{{\rm Li}^+}^{\rm ss} $. In conclusion, {\em{the  $\delta$ parameter should be carefully estimated prior to conducting any simulation}}.

\begin{figure} [t]
\centering
\includegraphics[width=0.75\columnwidth,keepaspectratio]{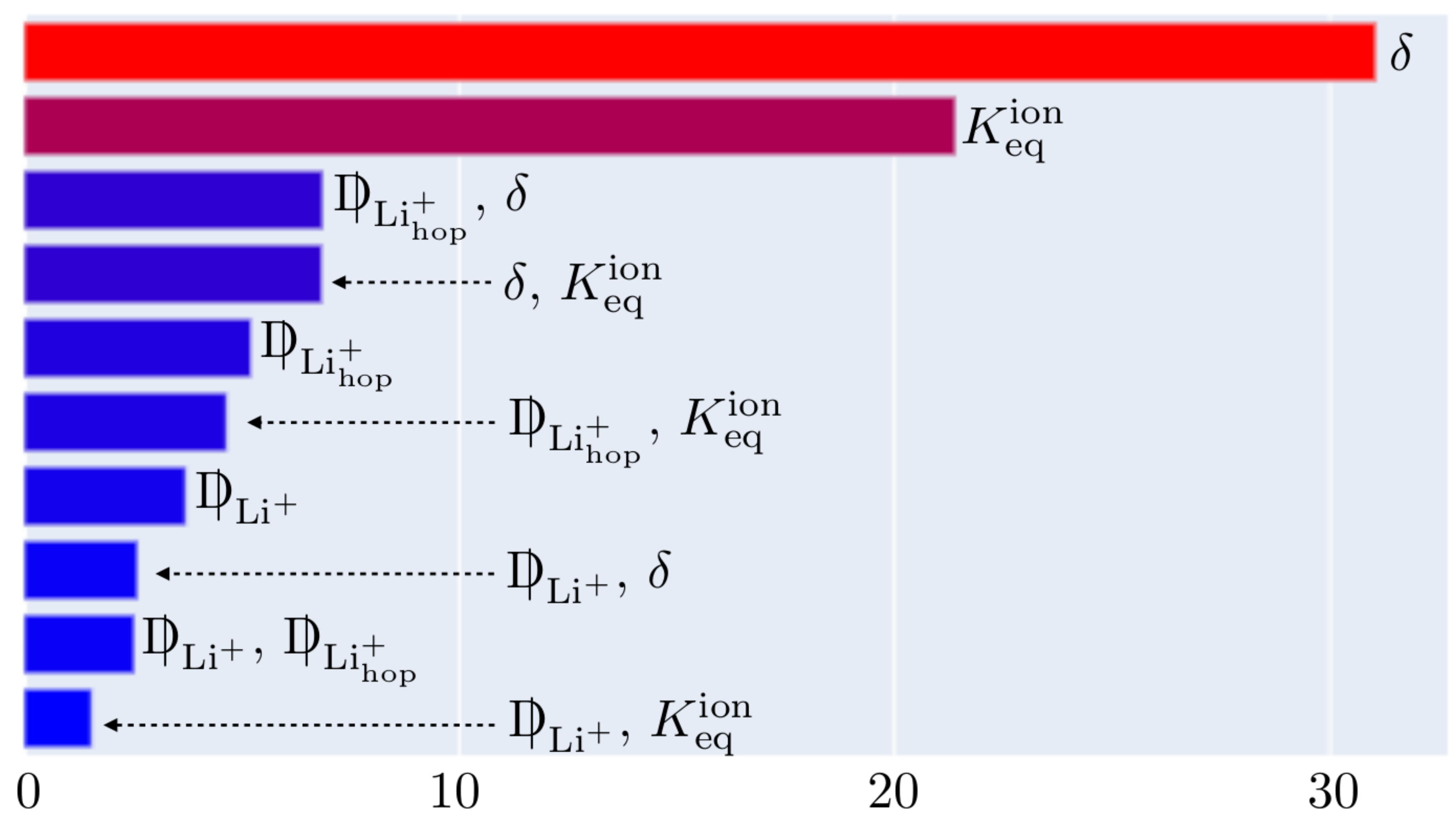}
 \caption{\em{The total Sobol indices for $b$.}}
 \label{b_sa}
 \end{figure}

Next, we study in Fig.~\ref{b_sa} the effect of the input parameters $\delta$,  $K_{\rm eq}^{\rm ion}$, $\diffusivity_{{\rm Li}^+_{\rm hop}}$, and $\diffusivity_{{\rm Li}^+}$ on the variability of the coefficient $b$ in eq. \eqref{eq:b_ss}. 
Once again majority of the variability is due to the variability in $\delta$. The variability of the parameter $K_{\rm eq}^{\rm ion}$ alone has more effect than the individual variability of each $\diffusivity_{{\rm Li}^+_{\rm hop}}$, and $\diffusivity_{{\rm Li}^+}$, while the interaction between  $\delta$ with each $\diffusivity_{{\rm Li}^+_{\rm hop}}$ and $K_{\rm eq}^{\rm ion}$ comes in the third place. These results suggest again that {\em $\delta$ is the most sensitive parameter, even to the electric potential coefficient $b$.}

\section{Conclusion}
\label{sec:conclusions}

In this note we thoroughly investigated the response of a novel model of solid electrolyte in terms of a major quantities of interest such as interface currents, the electric potential, fluxes and concentrations profiles. 
Governing equations, derived from continuity equations supplied with thermodynamically derived constitutive laws, have been either solved numerically via the finite element method or analytically in seeking for the steady state solution. The transient response shows a non-uniform profile of ionic concentrations, with gradients that attenuate in time towards a uniform ionic distribution at steady state. Depending upon the material parameters and initial conditions, the time required to complete the transient phase can be extremely long; in other words, the transient response might in fact be the only relevant behavior. This observation suggested us investigate the model sensitivity on the material parameters, with the aim of identifying the ones that contribute the most to the response. It turned out that the most relevant parameter is the fraction of Li that resides in equilibrium in the mobile state,
which can be accurately estimated according to \cite{RaijmakersEtAlEA2020}.

As proved in the companion paper \cite{CabrasEtAl2021b}, an accurate model for the solid electrolyte allows to reproduce well even the overall behavior of all solid state batteries. The quest of a complete understanding of the chemo-mechanics of SSBs and their still unresolved issues calls for a realistic description of the microstructure of composite cells. It is the target of our current research.


\bibliographystyle{jabbrv_unsrt}

\bibliography{/Users/albertosalvadori/Bibliography/Bibliography}

\appendix

\section{Equilibrium conditions for the chemical reactions}
\label{app:eqkin}

Equilibrium conditions for the chemical reactions \eqref{eq:IonizationReaction1} and \eqref{eq:IonizationReaction2} can be achieved from thermodynamics, as well. Imposing a vanishing affinity for \eqref{eq:IonizationReaction2}, for instance, leads to
\begin{equation}
\label{eq:affinity}
A^{y} = \elchempot_{{\rm Li}^+_{\rm hop}} - \elchempot_{{\rm Li}^+} = \mu_{{\rm Li}^+_{\rm hop}} - \mu_{{\rm Li}^+} = 0
\end{equation}
in view of definition \eqref{eq:electrochemicalpotential} since hopping and interstitial flows share the same electric potential. For ideal solutions, replacing eq. \eqref{eq:chempotidealnosat} into eq. \eqref{eq:affinity}, it comes out 
\begin{equation}
\label{eq:affinity2}
 \mu^0_{{\rm Li}^+_{\rm hop}} - \mu^0_{{\rm Li}^+} = RT \, \log \frac{ c^{\rm eq}_{{\rm Li}^+_{\rm hop}} }{ c^{\rm eq}_{{\rm Li}^+} } = RT \, \log K_{\rm eq}^{\rm hop} 
\end{equation}
by setting $y=0$ into eq. \eqref{eq:mass_action_hop}. Eq. relates $K_{\rm eq}^{\rm hop} $ to the negative of the Gibbs free energy change $\mu^0_{{\rm Li}^+_{\rm hop}} - \mu^0_{{\rm Li}^+} $.
The thermodynamic restriction
$$
y  \, A^{y}     \le 0
$$
is satisfied using Eq.(\ref{eq:mass_action_hop}). 
The affinity and the reaction rate can be restated as:
\begin{eqnarray*}
A^y
&=&   
RT \ln \left[ \frac{ \theta_{{{\rm Li}^+_{\rm hop}}} }{ 1-\theta_{{{\rm Li}^+_{\rm hop}}} } \; \frac{1- \theta_{{{\rm Li}^+}} }{ \theta_{{{\rm Li}^+}} } \,  \frac{1}{ K_{\rm eq}^{\rm hop}  } \right] 
     \; ,
\\
y
&=&
k_b^{\rm hop}
\left\{
-  \frac{ \theta_{{{\rm Li}^+_{\rm hop}}} }{ 1-\theta_{{{\rm Li}^+_{\rm hop}}} }
+
 \frac{ \theta_{{{\rm Li}^+}} }{ 1-\theta_{{{\rm Li}^+}} } \;   K_{\rm eq}^{\rm hop} 
\right\}
     \; .
\end{eqnarray*}
If $y > 0$ then 
$$
K_{\rm eq}^{\rm hop}   >   \frac{ \theta_{{{\rm Li}^+_{\rm hop}}} }{ 1-\theta_{{{\rm Li}^+_{\rm hop}}} } \; \frac{1- \theta_{{{\rm Li}^+}} }{ \theta_{{{\rm Li}^+}} }
$$
and in turn $A^y <0 $. Viceversa if $y  < 0$ then $A^y >0 $.

\end{document}